\documentclass[letterpaper,12pt]{article}
\usepackage[utf8]{inputenc}
\usepackage[margin=0.75in]{geometry}
\usepackage{mathtools}
\usepackage{amsmath}
\usepackage{placeins}
\usepackage{apacite}
\usepackage{color}
\usepackage{natbib}
\usepackage{bm}
\usepackage{dsfont}
\usepackage{array,float,booktabs}
\usepackage{sectsty}
\usepackage{longtable}
\usepackage[onehalfspacing]{setspace}

\begin{document}

\title{Conformational variability of loops in the SARS-CoV-2 spike protein}

\author{Samuel W.K. Wong\footnote{ Address for correspondence: Department of Statistics and Actuarial Science, University of Waterloo, Waterloo, ON, Canada.  E-mail: samuel.wong@uwaterloo.ca}~~and Zongjun Liu \\
	Department of Statistics and Actuarial Science, University of Waterloo 
}
\date{October 5, 2021}

\maketitle

\begin{abstract}
The SARS-CoV-2 spike (S) protein facilitates viral infection, and has been the focus of many structure determination efforts. Its flexible loop regions are known to be involved in protein binding and may adopt multiple conformations.  This paper identifies the S protein loops and studies their conformational variability based on the available Protein Data Bank (PDB) structures. While most loops had essentially one stable conformation, 17 of 44 loop regions were observed to be structurally variable with multiple substantively distinct conformations based on a cluster analysis. Loop modeling methods were then applied to the S protein loop targets, and the prediction accuracies discussed in relation to the characteristics of the conformational clusters identified. Loops with multiple conformations were found to be challenging to model based on a single structural template.

\textit{Key words and phrases:}  COVID-19, loop modeling, conformational ensembles, decoy selection, sequence variants, protein structure prediction

%\textit{Running title:} Analysis of COVID-19 spike protein loops

%The authors declare no conflict of interest.
\end{abstract}

\section{Introduction}

The COVID-19 disease is caused by the SARS-CoV-2 strain of coronavirus and its continued spread remains a concern since the first reported infections in late 2019 \citep{zhu2020novel}.  The SARS-CoV-2 viral genome encodes for four main structural proteins:  spike, envelope, membrane, and nucleocapsid \citep{jiang2020neutralizing}.  The spike (S) protein is of particular importance as it facilitates viral entry into host cells via its receptor binding domain (RBD), which recognizes human angiotensin-converting enzyme 2 \citep[ACE2,][]{shang2020structural}. Current vaccines being administered \citep[e.g.,][]{polack2020safety}  achieve efficacy against SARS-CoV-2 by enabling the human body to produce a modified version of its S protein; this in turn induces the production of neutralizing antibodies against the disease \citep{sewell2020covid}. 

Towards the development of such therapeutic interventions, many structure determination efforts have focused on the S protein, with the first standalone experimental structure of the full-length S protein obtained via cryo-electron microscopy in mid-February 2020 \citep{wrapp2020cryo}.  Soon thereafter, the structure of the S protein RBD bound in a complex with ACE2 was also determined \citep{lan2020structure}.  As of January 13th, 2021, there were 203 structures deposited in the Protein Data Bank \citep[PDB,][]{berman2000protein} associated with the SARS-CoV-2 S protein.  These include studies of the standalone S protein \citep[e.g.,][]{cai2020distinct}, the S protein interacting with potential antibodies \citep[e.g.,][]{shi2020human,schoof2020ultrapotent}, and the S protein interacting with various forms of ACE2  \citep[e.g.,][]{guo2021engineered}.  Finally, with the emergence of S protein sequence variants, structures corresponding to mutations are also being studied, with D614G being a common example \citep{yurkovetskiy2020structural}. While individual PDB structures generally provide static snapshots of protein conformations, it is well-known that proteins exhibit dynamic movement \citep{mittermaier2006new,henzler2007dynamic}. The local dynamics of atoms and residues are partially depicted via crystallographic B-factors \citep{schneider2014local}. Larger motions are also possible: for the SARS-CoV-2 S protein, a well-documented example is the ability of its RBD to adopt  `up' (or open) and `down' (or closed) states, where the `up' state is the conformation capable of binding to ACE2 \citep{wrapp2020cryo}. Overall then, the PDB is a rich source of data for examining the conformational variability of the S protein, given the number of times its structure has been solved experimentally.

This paper focuses on the loop conformations of the S protein.  Protein loops are the flexible connecting regions between regular secondary structures, and are where protein disorder is most likely to occur \citep{linding2003protein}. This greater disordered nature of loops may be manifest in a PDB structure via missing atomic coordinates or atoms with high B-factors \citep{shehu2006modeling}. Accurate structure prediction for loops is both challenging and necessary, to construct useful models for downstream therapeutic applications \citep{muhammed2019homology}. Loops are of particular importance as they are often associated with protein function, such as providing binding recognition sites and facilitating protein--protein interactions  \citep{espadaler2006identification}. For example, an extended loop of the SARS-CoV-2 S protein RBD interacts directly with loops of ACE2, as evidenced by the PDB structure of the RBD-ACE2 complex \citep{yan2020structural}.  Dynamic structural changes can occur both in larger regions of a protein (e.g., the SARS-CoV-2 RBD), as well as in individual loops adopting conformational rearrangements to carry out protein function in accordance with their environment \citep{papaleo2016role}.  Thus, when a protein has been solved many times in the PDB, we may be able to observe distinct conformations among some of its loops, given their potential for disorder and structural variability. In particular for the SARS-CoV-2 S protein, the PDB also documents sequence variants arising from mutations to some of its loop regions \citep[e.g.,][]{zhang2021structural}, and the possible structural impacts of mutations can also be studied more broadly via computational methods \citep{chen2020mutations,sedova2020coronavirus3d,wong2020assessing}. Mutations to the S protein are especially of concern as they can lead to more infectious variants of SARS-CoV-2 \citep{LiQ}.

The task of structure prediction for flexible loops with multiple distinct conformations has been found to be more challenging than for rigid or inflexible ones \citep{marks2018predicting}.  Most loop prediction methods are designed to identify the most likely conformation, e.g., with the lowest potential energy \citep{soto2008loop,stein2013improvements,liang2014leap,tang2014fast,wong2017fast,marks2017sphinx}.  Such methods are typically trained on loop sets where a single conformation for each loop is taken from the PDB and assumed to represent the ground truth \citep{fiser2000modeling}, and thus tend to be more successful at accurately predicting inflexible loops with one `correct' solution. Accuracy is typically measured by computing the root-mean-squared deviation (RMSD) of the backbone atoms from the predicted loop conformation to the corresponding one in the PDB. In order to study loops that can adopt multiple conformations, prediction methods might instead be applied to generate an ensemble of decoys, which often involves a combination of sampling and scoring steps \citep{barozet2021protein}.  Then, the success of different methods could be assessed on the basis of whether their generated ensembles include decoys that are close to each of the known conformations \citep{marks2018predicting}. For the SARS-CoV-2 S protein, this kind of assessment is a good test on the ability of current methods to explore a range of likely conformations, especially if further mutations were to occur in the flexible loop regions.  

These considerations motivate the main contributions of this paper.  First, we identify the loop regions and sequence variants from the known PDB structures of the SARS-CoV-2 S protein, and use cluster analysis to classify each loop according to whether it has been observed to adopt multiple distinct conformations or a single conformation only. Second, we apply four current loop prediction methods on the identified loop regions, to generate ensembles of decoys for each one. Third, we discuss the results of these methods and the effectiveness of their application to modeling the loops of the S protein, along with the insights gained via our analyses.  

\section{Materials and Methods}

\subsection{Data preparation and selection of loop targets} \label{sec:prep}

The 3-D structures of the SARS-CoV-2 S protein were downloaded from the PDB at the RCSB website (\url{https://rcsb.org}) on January 13th, 2021, by navigating to the page in the `COVID-19 coronavirus resources' section entitled `Spike proteins and receptor binding domains'. We extracted the S protein structures that are not bound to other molecules and have sequence length greater than 1000.  This facilitates study of the S protein loop conformations within the context of a (mostly) full-length S protein structure, while without explicit interaction with other proteins. A total of 63 S protein PDB structures satisfied these criteria, most of which are provided as S protein trimers. We treated each chain as an individual sample and thus extracted a total of 193 S protein chains. Some realignments of the corresponding amino acid sequences were required in order to keep the residue numbers consistent across all chains; this was accomplished with the ClustalO service in Jalview \citep{waterhouse2009jalview}.

For each S protein chain, we first used DSSP \citep{Kabsch1983} to determine the secondary structure classification of each residue.  The 8-state DSSP classification was reduced to the traditional three types of helix (H), sheet (E), and coil (C) following the conventions in the SPIDER3 \citep{heffernan2017capturing} secondary structure prediction method: we map DSSP's ``G'', ``H'', and ``I'' to H; ``E'' and ``B'' to E; the remaining three states are mapped to C.  Due to structural variability, the classified type (H, E, or C) for a given residue position may not always agree among the 193 S protein chains. Thus, we define a loop region for our study as follows: a segment of five or more consecutive residues where over 50\% of the protein chains at each position are classified as type C.  Further, if two such segments are separated by only one E or H type residue (i.e., where less than 50\% of the chains are type C at that position), we treat the two combined segments (including that connecting residue) as a single loop region.

With the starting and ending positions of loops defined in this manner, we check for the presence of sequence variants in each loop region among the S protein chains. If multiple distinct residue sequences are observed for a loop region, we shall treat each unique sequence separately for further analysis. This allows us to document the possible impact of mutations on the loop conformations.  Thus, we shall say that a loop instance consists of its starting and ending positions together with its unique residue sequence. We then consider the structural variability of each loop instance. To account for the potential disordered nature and structural uncertainties of loops, we extract both the atomic coordinates and B-factors from the PDB chains. Taking all chains that have no missing coordinates or B-factors within the loop residues, we compute their pairwise RMSD matrix based on the loop's backbone (N, C$_\alpha$, C, and O) atoms. The RMSD calculation is applied after the backbone atoms of the loop residues for each pair are optimally superimposed using the Kabsch algorithm \citep{kabsch1976solution}. This is the `local RMSD' \citep{choi2010fread,karami2019dareus} that compares the loop region only, and so is not sensitive to orientation differences in the rest of the structure.  Based on that distance matrix, we apply hierarchical clustering with average linkage  \citep[UPGMA,][]{sokal1958statistical} and a distance cutoff of 1.5 \AA~\citep{marks2018predicting} to form initial clusters of loop conformations.

Following, we incorporate B-factors to ensure that the clusters formed are statistically distinct. Recall that the B-factor can be expressed in terms of the mean-square amplitude of atomic oscillations $u^2$ around their measured positions: $B = 8 \pi^2 \left<u^2\right>$. Using an isotropic Gaussian approximation for the corresponding coordinate uncertainties, we can determine whether the difference in backbone coordinates between a loop pair is significantly different with 95\% confidence (see Appendix A for details). If none of the chains in one cluster are significantly different from any chains in another cluster, we merge them into a single cluster. Clusters composed entirely of chains with poor structure resolution ($>3$ \AA) after this step are removed from further analysis as the atomic coordinates are unlikely to be sufficiently reliable for making detailed structural comparisons. Each remaining cluster then represents a distinct group of S protein chains which have a similar conformation for that loop instance.  We consider a loop instance to have multiple distinct conformations if this analysis results in two or more such clusters of conformations; otherwise, we say that loop instance essentially adopts only a single conformation. We select a representative from each cluster by taking the chain with resolution $\le 3$ \AA~that is closest to the geometric centroid of the cluster.

Our full list of S protein loop targets for study thus consists of all the cluster representatives obtained from the above steps.

\subsection{Loop modeling methods} \label{sec:loopmodel}

To study the conformational variability of the identified S protein loop targets, we make use of several loop modeling methods. We focus on methods that incorporate sampling-based techniques for loop construction, which are suitable for stochastically generating an ensemble of decoys that represent plausible conformations for a loop.  We include Rosetta's next-generation KIC  (NGK) algorithm \citep{stein2013improvements}, the DiSGro algorithm \citep{tang2014fast}, and the PETALS algorithm \citep{wong2017fast}, which are \emph{ab initio} methods that explore the conformational space with the guidance of an energy or scoring function; these do not directly make use any structure templates of known loop conformations.  We also include the Sphinx algorithm \citep{marks2017sphinx}, which is a hybrid method that begins with loop structure fragments obtained from sequence alignment and then completes the loop construction by \emph{ab initio} sampling.

Using each of the methods, we generate an ensemble of 500 decoys for each loop target. The input (or template) structure is the loop target's representative PDB chain, prepared by removing the coordinates of the loop residues: following loop modeling conventions, we treat the backbone atoms from the starting residue's C atom to the ending residue's C$_\alpha$ atom as unknown. The generated decoys are compared with the loop structures from each known conformation for that loop region. The backbone RMSD is used to assess the accuracy of the decoys. Two types of RMSDs are calculated, as in \citet{choi2010fread}: local RMSD (which superimposes the backbone of the loop residues, as in section 2.1) and global RMSD, which superimposes the backbone atoms of the two residues on either side of the loop (rather than the backbone of the loop residues themselves) prior to the calculation. Global RMSD, as often reported in loop modeling studies, also considers the decoy's orientation to the rest of the structure. For loop regions with multiple conformations or mutations, decoy generation is carried out multiple times, once using each representative PDB as input; taken together, we may thus assess whether decoys generated from different PDB inputs have good coverage of the conformational space for that loop region.

The scoring function associated with each method provides a ranking of its 500 generated decoys for a loop target. Thus, it is of interest to assess how well each method's top-ranking decoys can predict the possible conformations of the loop region. We use three RMSD statistics for this purpose: (a) lowest RMSD among the 500 decoys, (b) RMSD of the top-ranked decoy, (c) lowest RMSD among the top-five ranked decoys.  The first RMSD statistic evaluates the method according to its ability to construct native-like conformations, without regard to whether its scoring function can select the best prediction.  The second RMSD statistic corresponds to typical loop modeling assessment, where the top-ranked decoy is selected as the prediction.  However, this approach of selecting a single prediction would be less informative if the loop region has multiple conformations.  Thus, we also use the third RMSD statistic: by selecting multiple (i.e., the top five) decoys, we can examine whether these top-ranking decoys are structurally distinct and accurately represent the different known conformations.  

We briefly describe how each of the loop modeling methods is run. The NGK algorithm \citep{stein2013improvements} is included in the Rosetta protein modeling suite (available at  \\ \url{https://www.rosettacommons.org/}), and we used the version provided in Rosetta release 2020.50 on December 18, 2020.  NGK improves on a previous kinematic closure method, which consists of local conformational sampling and Monte Carlo minimization steps performed over two (coarse and full-atom) stages.  The program outputs the lowest energy loop structure found in each run, and so to obtain the desired ensemble of decoys we ran the program 500 times, following the recommended settings in the online guide (\url{https://guybrush.ucsf.edu/benchmarks/benchmarks/loop\_modeling}).  The DiSGro algorithm \citep{tang2014fast} uses a distance-guided sequential chain-growth method to stochastically sample loop structures. We ran the authors' program to generate 100,000 conformations for the best possible coverage of the conformational space, then used their scoring function to select the 500 decoys with the lowest energy.  The PETALS algorithm \citep{wong2017fast} uses a sequence of propagation and filtering steps to explore the conformational space and locate low-energy structures.  We ran the authors' program with 60,000 seeds and outputted 30,000 decoys, then used an updated scoring function to select the 500 top-ranked decoys, see Appendix B for details.  The Sphinx algorithm \citep{marks2017sphinx} begins by searching a database for suitable fragments according to loop sequence alignments;  loop decoy backbones are then constructed by sampling and ranked with a coarse-grained energy function, after which side chains are added and SOAP-Loop \citep{dong2013optimized} is used to obtain the final ranking of decoys.  Sphinx is hosted on the SAbPred server \citep{dunbar2016sabpred}, for which we automated the loop target submissions and used the ``general protein'' option; no PDB blacklist was necessary as the fragment database had not yet been updated to contain any COVID-19 S protein structures.

\section{Results and discussion}

\subsection{Loop targets of the SARS-CoV-2 S protein}\label{sec:targets}

Applying the procedures in section \ref{sec:prep} to the 193 standalone S protein chains, a total of 44 loop regions were identified in the SARS-CoV-2 S protein.  Their starting and ending residue positions are listed in the first column of Table \ref{tab:seqlist}.  Thirty-two of the 44 loops lie within the S1 subunit, with 13 in the N-terminal domain and 11 in the RBD; e.g., loops 475--487 and 495--506 have been previously noted to form contacts with ACE2 during binding \citep{ali2020dynamics}.  Loop sequences are shown in the second column of Table \ref{tab:seqlist}. There are five loop regions with sequence variants in the PDB: 380--394, 410--416, 600--608, 614--620, and 891--897.  For these loop regions, the most common variant in the PDB is shown first, followed by the other variants which have their mutated residue indicated in bold.  The mutation that has received the most attention thus far is D614G \citep[e.g.,][]{yurkovetskiy2020structural,grubaugh2020making,zhang2020sars}. In total there are 50 loop instances, i.e., the combination of a loop's residue positions and unique amino acid sequence. The third column of Table \ref{tab:seqlist} shows the number of PDB chains that contain a complete backbone (i.e., atomic coordinates and B-factors) for each loop instance.

\begin{table}[!htbp]
	\caption{SARS-CoV-2 S protein loops. The first column shows the starting and ending positions of each identified loop region. The second column shows the loop sequences; if there are sequence variants in the PDB, the most common variant is listed first, and other variants have their mutated residues marked in bold.  The number of PDB chains containing that loop instance are shown in the third column.  The rightmost column lists the representative PDB chains for each loop instance; if a loop instance has multiple conformations, each chain listed corresponds to one distinct conformation (cluster). The number of PDB chains represented by each cluster is shown in parentheses; these may not sum up to the third column since clusters with poor structure resolution (all chains $>3$\AA) are omitted. \strut}
	\centering
		\scriptsize
	\begin{tabular}{lp{5cm}rl}
		\hline
		\textbf{Region} & \textbf{Sequence} & \textbf{\#Chains} & \textbf{Representative conformations} \\ 
		\hline
		14-27 & QCVNLTTRTQLPPA & 36 & 6zgeA(24), 7dddC(12) \\ 
		31-46 & SFTRGVYYPDKVFRSS & 185 & 7a4nB(185) \\ 
		56-60 & LPFFS & 185 & 6xr8A(185) \\ 
		66-83 & HAIHVSGTNGTKRFDNPV & 11 &  none (all PDBs $>3$\AA~resolution) \\ 
		108-116 & TTLDSKTQS & 169 & 6zoxB(169) \\ 
		130-140 & VCEFQFCNDPF & 168 & 6xluB(145), 7kdkC(5), 7kdlA(4) \\ 
		146-168 & HKNNKSWMESEFRVYSSANNCTF & 38 & 6zgiB(27), 7dddC(9) \\ 
		172-187 & SQPFLMDLEGKQGNFK & 52 & 7df3B(39), 6zp0B(12) \\ 
		210-222 & INLVRDLPQGFSA & 154 & 6vxxA(152) \\ 
		230-236 & PIGINIT & 185 & 6vxxA(185) \\ 
		245-263 & HRSYLTPGDSSSGWTAGAA & 26 & 6zgiB(24) \\ 
		280-284 & NENGT & 185 & 6x79B(185) \\ 
		304-310 & KSFTVEK & 185 & 7a4nB(185) \\ 
		320-324 & VQPTE & 185 & 6zoxC(181), 6xm3A(4) \\ 
		329-338 & FPNITNLCPF & 180 & 6x29A(155), 7kdlB(21) \\ 
		343-348 & NATRFA & 181 & 6zgeC(181) \\ 
		370-375 & NSASFS & 182 & 6vxxA(139), 6zgiC(42) \\ 
		380-394 & YGVSPTKLNDLCFTN & 170 & 7kdlC(164) \\ 
		& YGV\textbf{C}PTKLNDLCFTN & 12 & 6x79B(12) \\ 
		410-416 & IAPGQTG & 179 & 7kdkA(178) \\ 
		& IAP\textbf{C}QTG & 3 & 6zoxB(3) \\ 
		422-430 & NYKLPDDFT & 182 & 6xr8B(178), 6xm0B(2) \\ 
		438-451 & SNNLDSKVGGNYNY & 93 & 6xr8A(85), 7kdlB(4) \\ 
		454-472 & RLFRKSNLKPFERDISTEI & 96 & 6zgeC(95) \\ 
		475-487 & AGSTPCNGVEGFN & 92 & 7dddA(87), 6xm0B(1) \\ 
		495-506 & YGFQPTNGVGYQ & 124 & 6zp0A(118), 6xm0B(2), 7kdlB(3) \\ 
		517-523 & LLHAPAT & 168 & 6zoxA(163), 6xm0A(2), 6xm0B(1), 6xm3A(2) \\ 
		526-537 & GPKKSTNLVKNK & 181 & 7ad1B(26), 6x29B(154) \\ 
		555-564 & SNKKFLPFQQ & 185 & 7kdkC(185) \\ 
		578-583 & DPQTLE & 185 & 6zoxB(185) \\ 
		600-608 & PGTNTSNQV & 170 & 7kdlA(169) \\ 
		& PGTNTSN\textbf{E}V & 12 &  none (all PDBs $>3$\AA~resolution) \\ 
		614-620 & DVNCTEV & 103 & 6xm4C(98) \\ 
		& \textbf{G}VNCTEV & 42 & 7kdkA(42) \\ 
		& \textbf{N}VNCTEV & 6 & 7a4nB(6) \\ 
		624-641 & IHADQLTPTWRVYSTGSN & 26 & 6xm0B(18) \\ 
		656-663 & VNNSYECD & 185 & 7kdkB(185) \\ 
		697-710 & MSLGAENSVAYSNN & 185 & 6vxxB(185) \\ 
		783-816 & AQVKQIYKTPPIKDFGGFNFS... \mbox{~~}...QILPDPSKPSKRS & 144 & 6zp0C(142) \\ 
		825-836 & KVTLADAGFIKQ & 39 & 6xluB(2), 6xm3B(5), 6xm3C(1), 6zgiA(25) \\ 
		841-848 & LGDIAARD & 43 & 6xluC(6), 6xm4B(1), 6zgeB(20), 6xm3B(6), 7dddB(6) \\ 
		862-866 & PPLLT & 185 & 6zoxB(185) \\ 
		891-897 & GAALQIP & 176 & 7kdkB(176) \\ 
		& G\textbf{P}ALQIP & 9 & 7a4nB(9) \\ 
		908-913 & GIGVTQ & 185 & 7a4nB(185) \\ 
		968-976 & SNFGAISSV & 188 & 6zp0C(185), 6xraC(3) \\ 
		1033-1046 & VLGQSKRVDFCGKG & 188 & 7kdkA(188) \\ 
		1106-1112 & QRNFYEP & 188 & 7kdkC(188) \\ 
		1124-1132 & GNCDVVIGI & 188 & 6xm0A(185), 6xraC(3) \\ 
		1135-1141 & NTVYDPL & 161 & 7kdkB(158), 6xraC(3) \\ 
		\hline
		\end{tabular}
	\label{tab:seqlist}
\end{table}

\begin{figure}[ht]
	\centering
	\begin{tabular}{m{7.5cm}m{7.5cm}}
		\includegraphics[width=7.5cm]{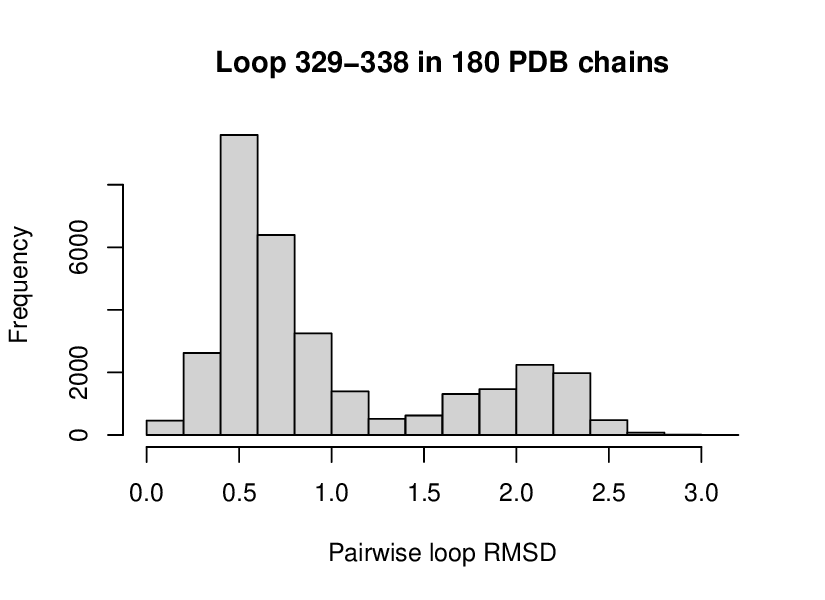} & 
		\includegraphics[width=7.5cm]{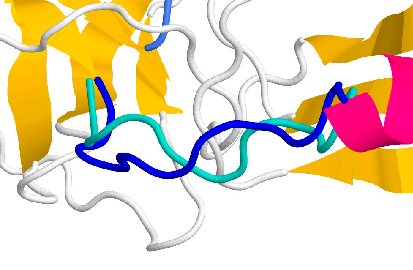}\\
		\includegraphics[width=7.5cm]{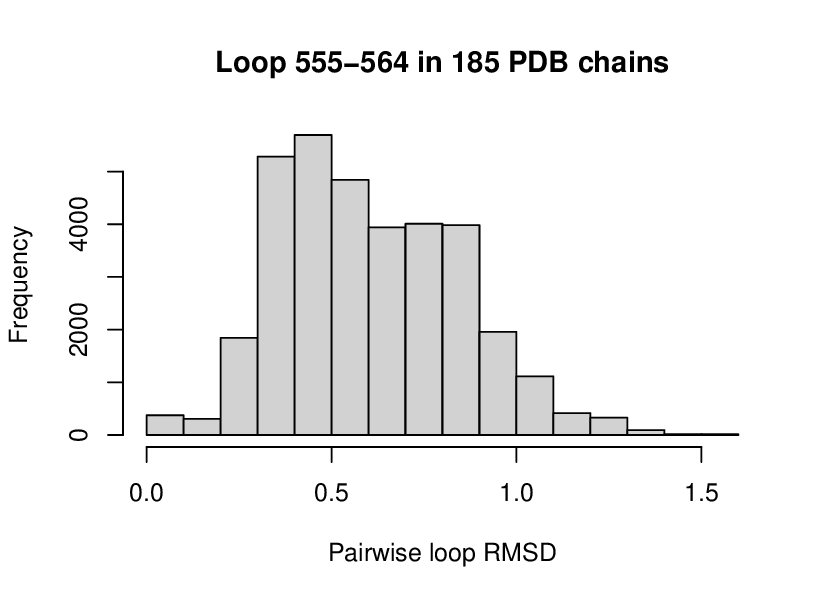} & 
		\includegraphics[width=7.5cm]{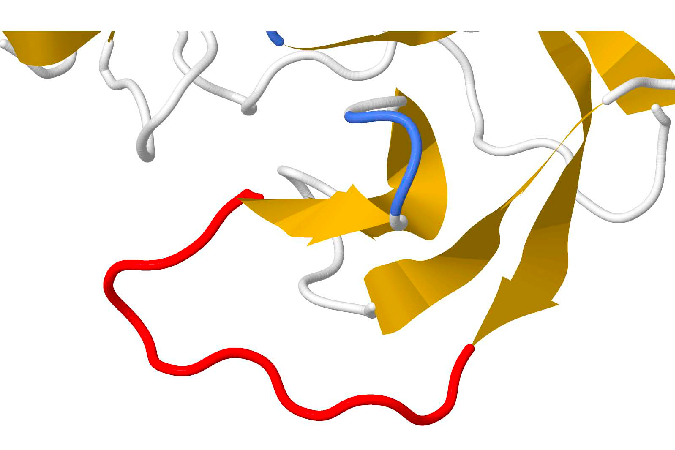}\\			
		
	\end{tabular}
	\caption{Two examples of SARS-CoV-2 S protein loops of length 10: 329--338 (top panels) and 555--564 (bottom panels).  The histograms (left panels) shows the pairwise RMSDs of the loop backbone among all S protein chains containing that loop: it can be seen that 329--338 exhibits higher structural variability than 555--564, due to the presence of two distinct clusters.  The right panels display close-ups of the representative loop conformations:  329--338 has two distinct conformations, colored in dark blue and turquoise; 555--564 has essentially one conformation, colored in red.}
	\label{fig:confs}
\end{figure}

The final column lists the representative PDB chains for each loop instance, obtained by the procedure for constructing clusters as described in section \ref{sec:prep}.  Thus, for example, there are 180 S protein chains that contain the loop at positions 329--338; clustering by pairwise RMSD identified two distinct conformations among structures with resolution $\le 3$\AA; 6x29A and 7kdkC were chosen to represent these clusters (which included 155 and 21 chains respectively), being the chains with resolution $\le 3$\AA~closest to the cluster centroids.  We illustrate the 329--338 loop example in the top panels of Figure \ref{fig:confs}:  a histogram of all pairwise RMSDs of the loop backbone (among the 180 S protein chains that contain this loop) is shown on the left, while a close-up of the part of the S protein chain containing the loop is shown on the right.  The histogram shows distinct peaks at pairwise loop RMSDs of 0.4--0.6 \AA~and 2.0--2.4 \AA, from which clustering identified the two distinct conformations colored dark blue and turquoise.  In contrast, the bottom panels of Figure \ref{fig:confs} show another length 10 loop region (555--564) but with little structural variability: the pairwise RMSDs do not exceed around 1.5 \AA~and clustering identified just one main conformation (colored in red). 

The initial hierarchical clustering step resulted in 137 clusters for the 50 loop instances. Based on the B-factor calculations, 17 of the 137 clusters did not have statistically distinct atomic coordinates compared to other clusters, and so merging these resulted in 120 clusters. All of the 17 clusters being merged had also failed to contain structures with sufficient resolution ($\le 3$~\AA). A further 45 of the 120 clusters contained no $\le 3$ \AA~structures, which led to two of the loop instances being omitted: 66-83 and 600-608 with the Q607E mutation. The final 75 clusters thus covered 48 loop instances; 17 of the 48 had multiple  distinct conformations (ranging from two to five).  By choosing the centroid of each cluster as its representative conformation, a diverse set of 41 different PDB chains with $\le 3$ \AA~resolution can be seen in Table \ref{tab:seqlist}. It should be noted that the exact number and composition of clusters will depend on the algorithm (i.e., cutoff and criterion) chosen.  Here using a cutoff of 1.5 \AA~with UPGMA, the average RMSD between members of different clusters will be at least 1.5 \AA. For example, if we used a cutoff of 1.5 \AA~with WPGMA \citep{sokal1958statistical} instead, 42 of the 50 loop instances maintain the same final clustering results; WPGMA would have found 82 representative conformations for the 48 loop instances.  Overall, we consider the clusters in Table \ref{tab:seqlist} to provide a fairly stable characterization of the structural variability present in these loops.

\begin{figure}[ht]
	\centering
		\includegraphics[width=7.5cm]{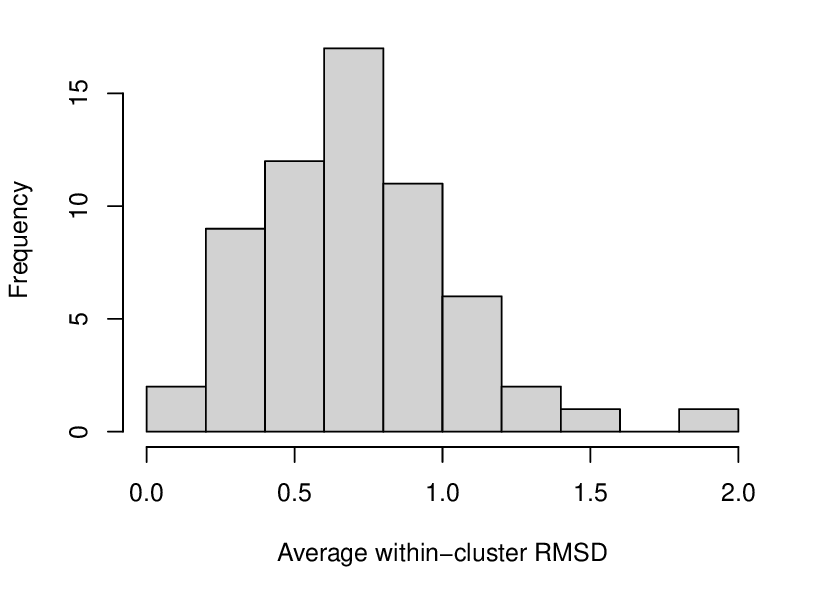}
	\caption{The amount of within-cluster variation for the 61 clusters defined by at least four chains and two distinct PDB codes. The breadth of movement observed within a cluster is measured by its average within-cluster RMSD; 36 of the clusters have an average between 0.5 to 1 \AA.}
	\label{fig:withinrmsd}
\end{figure}

The final 75 clusters in Table \ref{tab:seqlist} differ in their size and within-cluster variation. There were 4 singleton clusters (defined by a single chain only), and 61 clusters were defined by at least four chains and two distinct PDB codes (and often significantly more). These high chain counts per cluster enable more cluster statistics to be examined, compared to related studies, e.g., \citet{marks2018predicting} where clusters were defined by at most 5 chains (except in one case). Here, loop instances with multiple conformations tend have a dominant cluster that is defined by at least two-thirds of the available chains; the one exception is 841--848, which is also the most structurally variable loop with five distinct clusters. For each of the 61 well-represented clusters, we computed the average within-cluster RMSD (i.e., between all pairs of members in that cluster) as a measure of its breadth of movement, and a histogram is shown in Figure \ref{fig:withinrmsd}. The average breadth over all 61 clusters is 0.72 \AA. The list of clusters grouped according to their breadth $d$ is shown in Table \ref{tab:withinlist}, where 16 clusters are fairly tight with $d \le 0.5$ \AA, 36 clusters have $0.5 < d \le 1.0$, and the 10 loosest clusters have $d > 1.0$ \AA. It might be expected that shorter loops tend to form tighter clusters as they have a smaller conformational space; indeed, this pattern can be seen as the average loop length of clusters in these three groups are 6.5, 12.1, and 13.0 respectively. The larger clusters also tend to be tighter: the average cluster size in these three groups are 127, 108, and 49, respectively. However, we note that these are overall patterns only; for example, the cluster for the longest loop 783--816 is defined by 142 chains and has only a moderate $d = 0.81$.

\begin{table}[!htbp]
		\footnotesize	
	\centering
	\caption{Clusters grouped according to their breadth of movement $d$ as defined by their average within-cluster RMSDs. Each cluster is listed based on its representative conformation (Table \ref{tab:seqlist}) together with its starting and ending residues. The average loop length and size of clusters in the three groups are shown in the rightmost columns. \strut}
	\begin{tabular}{l>{\raggedright\arraybackslash}p{10cm}cc}
		\textbf{Breadth ($d$)} & \textbf{Clusters} & \textbf{Avg.~length} & \textbf{Avg.~size} \\
		\hline  
		$d \le 0.5$ \AA & 6xr8A\_56\_60, 6x79B\_280\_284, 7a4nB\_304\_310, 6zoxC\_320\_324, 6xm3A\_320\_324, 7kdkA\_614\_620, 7a4nB\_614\_620, 7kdkB\_656\_663, 7dddB\_841\_848, 6zoxB\_862\_866, 7kdkB\_891\_897, 7a4nB\_891\_897, 7a4nB\_908\_913, 6zp0C\_968\_976, 7kdkC\_1106\_1112 & 6.5 & 127\\
		\hline
$0.5 < d \le 1.0$ \AA &  6zgeA\_14\_27, 7a4nB\_31\_46, 6zoxB\_108\_116, 6xluB\_130\_140, 7kdlA\_130\_140, 7dddC\_146\_168, 7df3B\_172\_187, 6zp0B\_172\_187, 6vxxA\_230\_236, 6x29A\_329\_338, 7kdlB\_329\_338, 6zgeC\_343\_348, 6vxxA\_370\_375, 6zgiC\_370\_375, 7kdlC\_380\_394, 6x79B\_380\_394, 7kdkA\_410\_416, 6xr8B\_422\_430, 6xr8A\_438\_451, 6zgeC\_454\_472, 6zp0A\_495\_506, 7ad1B\_526\_537, 6x29B\_526\_537, 7kdkC\_555\_564, 6zoxB\_578\_583, 7kdlA\_600\_608, 6xm4C\_614\_620, 6xm0B\_624\_641, 6vxxB\_697\_710, 6zp0C\_783\_816, 6xm3B\_825\_836, 6zgiA\_825\_836, 6zgeB\_841\_848, 7kdkA\_1033\_1046, 6xm0A\_1124\_1132, 7kdkB\_1135\_1141 & 12.1 & 108\\
\hline
$d > 1.0$ \AA & 7dddC\_14\_27, 7kdkC\_130\_140, 6zgiB\_146\_168, 6vxxA\_210\_222, 6zgiB\_245\_263, 7kdlB\_438\_451, 7dddA\_475\_487, 6zoxA\_517\_523, 6xluC\_841\_848, 6xm3B\_841\_848  & 13.0 & 49\\
		\hline
	\end{tabular}
	\label{tab:withinlist}
\end{table}

It is well-known that the SARS-CoV-2 RBD as a whole can adopt an `up' or `down' conformational state \citep{wrapp2020cryo}.  Seven of the 17 loop instances with multiple conformations were located within the RBD. Notably, both 475--487 and 495--506 which interact with ACE2 are among these. Thus, we examined whether this higher propensity for multiple conformation loops within the RBD might be associated with the chains having an `up' or `down' RBD state, even when the S protein chain is considered in isolation. We took PDB 6zge \citep{wrobel2020sars}, where it is known that chain A has a `down' RBD and chain B has an `up' RBD.  Then, each of the 193 S protein chains was classified as `up' or `down' according to whether its backbone RMSD to 6zgeB or 6zgeA was smaller.  Based on this criterion, the loop at 370--375 has both distinct conformations coming from `down' RBD chains, while four other loops with two conformations (329--338, 422--430, 438--451, 475--487) indeed have one conformation associated with the `up' state and the other associated with the `down' state.  Of the two remaining loops, 495--506 has one conformation from an `down' RBD and two from an `up' RBD, while 517--523 has two conformations from each.  Overall then, five RBD loop regions have structures that do not vary significantly with the RBD state (370--375 and the four single conformation loops in the RBD), while the other six do potentially vary.

Five loop regions had sequence variants present in the PDB, each consisting of a single point mutation. All of these loop instances had only a single conformation. Taking the representative chain for each sequence variant listed in Table \ref{tab:seqlist}, we computed the local loop backbone RMSD between the representatives and  the results are shown in Table \ref{tab:mut}.  For example, for the loop region 380--394, the sequence variants are S and C at position 383, represented by 7kdlC and 6x79B respectively; these structures have backbone RMSD 0.54 \AA~computed on the loop residues.  For the loop 600--608, there were no high resolution PDB structures containing the Q607E mutation. Overall, these sequence variants do not have large impacts on the loop conformations with observed backbone differences all $<1$ \AA, such that the conformational space of these loop regions (including variants) could be represented by a single cluster.

\begin{table}[!htbp]
	\centering
	\caption{Backbone RMSDs between the PDB chains representing the different sequence variants, in loop regions where mutations are present.  Local RMSDs are computed on the loop residues.  The residues that differ between the sequence variants are highlighted in bold.}
	\begin{tabular}{llll}
	\textbf{Region} & \textbf{Sequence 1} & \textbf{Sequence 2} & \textbf{RMSD} \\
	\hline  
380-394 &  YGV\textbf{S}PTKLNDLCFTN & YGV\textbf{C}PTKLNDLCFTN & 0.54 \\
410-416 & IAP\textbf{G}QTG & IAP\textbf{C}QTG &  0.40 \\
614-620 & \textbf{D}VNCTEV & \textbf{G}VNCTEV & 0.67 \\
614-620 & \textbf{D}VNCTEV & \textbf{N}VNCTEV & 0.62 \\
614-620 & \textbf{G}VNCTEV & \textbf{N}VNCTEV & 0.51 \\
891-897 & G\textbf{A}ALQIP & G\textbf{P}ALQIP &  0.23 \\
\hline	
	\hline
	\end{tabular}
\label{tab:mut}
\end{table}

Three of the loop targets were omitted from consideration for loop modeling, as all of their PDB chains were missing a residue immediately next to the loop: 14--27 (both conformations missing residue 13), 614--620 with the D614G and D614N mutations (both missing residue 621).  Thus the loop modeling methods were applied to a total of 71  targets.

\subsection{Loop modeling results}

The four methods described in section \ref{sec:loopmodel} were applied to model the conformations of the 71 loop targets identified in section \ref{sec:targets}. Of these, 66 targets could be run successfully using all four methods.  NGK and PETALS completed decoy generation for all 71 targets, while DiSGro completed 68 targets and Sphinx completed 66 targets. We focus the discussion on the results of the 66 loop targets for which all the methods could successfully generate decoys; the 5 remaining cases are discussed briefly at the end.

First, we assess the ability of methods to predict a correct loop structure. We define this loop prediction accuracy by calculating the RMSD to the closest loop structure among all chains containing that loop instance. Thus for this task, a good prediction can be close to \textit{any} cluster member among \textit{any} of the loop's known conformations (clusters), which accounts for the possible within-cluster variation (Figure \ref{fig:withinrmsd}) and treats loop structures in all the chains as an equi-energetic ensemble. Loop targets representing regions with multiple conformations can score well by this definition as long as a method can predict any one of the known conformations. For example, there are three targets for the loop 130--140 corresponding to its three conformations, represented by 6xluB, 7kdkC, and 7kdlA; decoys generated using 6xluB as input are compared to loop structures in all 154 chains of the three clusters combined, and likewise for 7kdkC and 7kdlA. We categorized the targets according to whether they belong to loop instances with multiple conformations or not; these categories are denoted as `Multiple conf.' and `Single conf.' in Table \ref{tab:rmsdpred}, containing 40 and 26 loop targets respectively. Table \ref{tab:rmsdpred} displays the three RMSD statistics described in the Materials and Methods section --  lowest RMSD among the 500 decoys, RMSD of the top-ranked decoy, and lowest RMSD among the top-five ranked decoys -- using both local and global RMSD calculations and averaged over the loop targets for each method. On average, all four methods can generate decoys at $<$1 \AA~local RMSD and $<$1.5 \AA~global RMSD from a correct structure. However, it remains difficult to correctly rank the generated decoys, with the RMSDs of the top-ranked decoy often substantially higher than the best decoy available. When each method is allowed to choose five decoys, then it is more likely that at least one of the five is close to a correct structure; e.g., NGK's average accuracy improves from 2.31 to 1.60 \AA~(global RMSD).  Further, the difficulty of the loop prediction task tends to vary by target category: for all four methods, the average top decoy RMSD for loops with multiple conformations are higher than for single conformation loops, whether considering local or global RMSDs.

\begin{table}[ht]
	\caption{RMSD metrics for assessing the loop prediction accuracy of the four methods.  The loop backbone RMSDs shown are averaged over single conformation targets ($n = 26$), multiple conformation targets ($n = 40$), and all targets ($n = 66$). The columns `Min.', `Top', and `Top-5' refer respectively to the lowest RMSD among the 500 decoys, RMSD of the top-ranked decoy, and lowest RMSD among the top-five ranked decoys. Prediction accuracy is defined as the RMSD to the closest loop structure among all chains containing that loop instance.}
	\centering
	\begin{tabular}{llcccccc}
		& & \multicolumn{3}{c}{Local RMSD} & \multicolumn{3}{c}{Global RMSD}   \\ \cmidrule(lr){3-5} \cmidrule(lr){6-8}
		\textbf{Method} & \textbf{Target category} & \textbf{Min.} & \textbf{Top} & \textbf{Top-5} & \textbf{Min.} & \textbf{Top} & \textbf{Top-5} \\ 
		\hline
		& Single conf. & 0.76 & 1.81 & 1.28 & 0.97 & 2.66 & 1.73 \\ 
		DiSGro & Multiple conf. & 0.96 & 1.95 & 1.56 & 1.47 & 3.60 & 2.95 \\ 
		& All & 0.88 & 1.90 & 1.45 & 1.27 & 3.23 & 2.47 \\ 
		\hline
		& Single conf. & 0.42 & 1.06 & 0.85 & 0.58 & 1.93 & 1.62 \\ 
		NGK & Multiple conf. & 0.66 & 1.42 & 1.08 & 1.07 & 2.55 & 1.59 \\ 
		& All & 0.56 & 1.28 & 0.99 & 0.87 & 2.31 & 1.60 \\ 
		\hline
		& Single conf. & 0.68 & 1.24 & 0.98 & 0.98 & 2.06 & 1.51 \\ 
		PETALS & Multiple conf. & 0.85 & 1.58 & 1.33 & 1.42 & 3.00 & 2.32 \\ 
		& All & 0.78 & 1.44 & 1.19 & 1.25 & 2.63 & 2.00 \\ 
		\hline
		& Single conf. & 0.64 & 1.49 & 1.15 & 1.11 & 2.75 & 2.09 \\ 
		Sphinx & Multiple conf. & 0.74 & 1.77 & 1.31 & 1.34 & 3.53 & 2.46 \\ 
		& All & 0.70 & 1.66 & 1.25 & 1.25 & 3.22 & 2.31 \\ 
		\hline
	\end{tabular}
	\label{tab:rmsdpred}
\end{table}

\begin{figure}[!htbp]
	\centering
	\includegraphics[width=18.5cm]{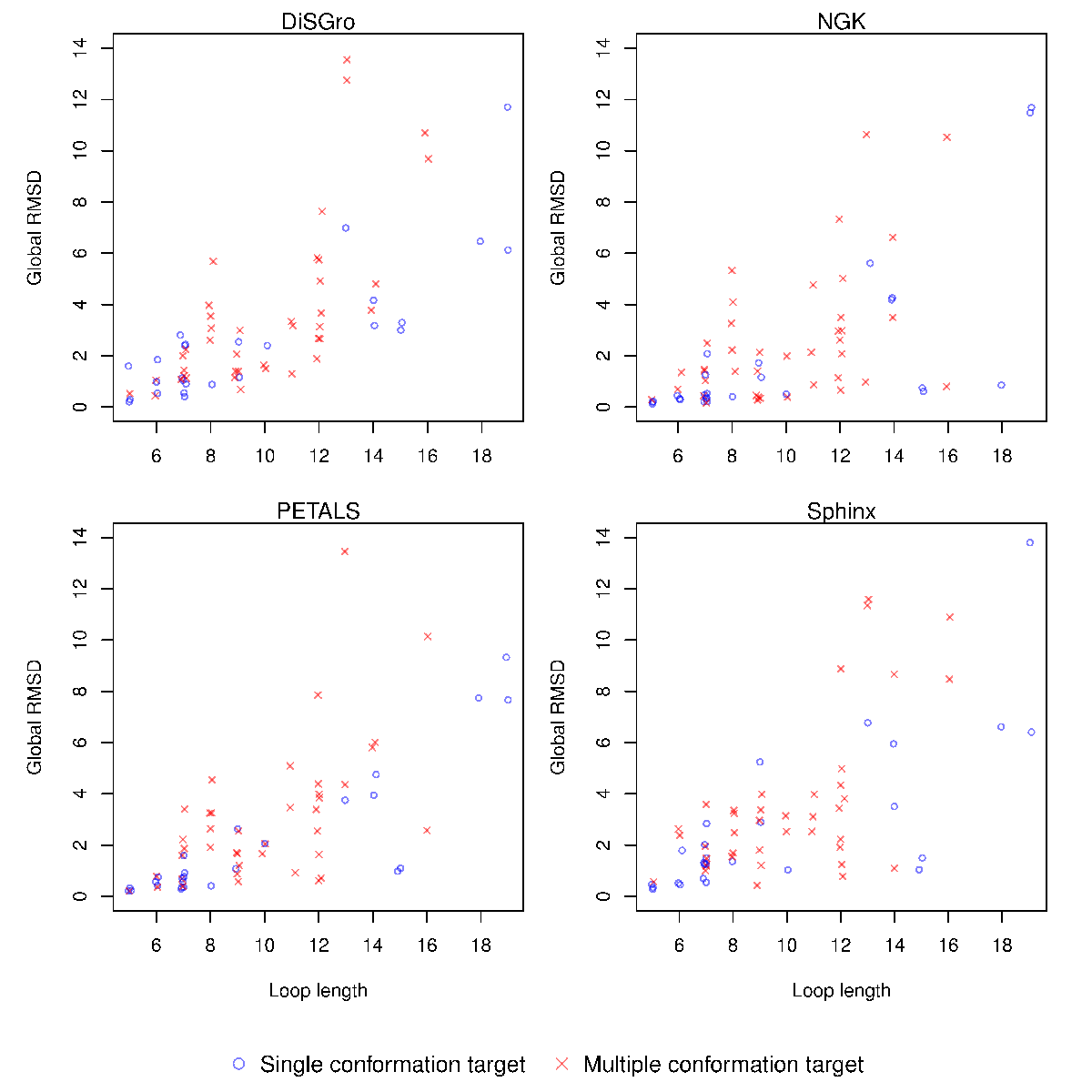}
	\caption{Loop prediction accuracy for each of the four methods, visualized by plotting the global RMSD of the top decoy vs.~loop length. Prediction difficulty increases with loop length, with methods consistently achieving $<$2 \AA~RMSD only for the shortest loops ($\le 6$ residues). The hardest targets for a given loop length tend to be those from multiple conformations, especially for the two most accurate methods (NGK and PETALS). Slight jitter is added  along the x-axis to the points  for readability.}
	\label{fig:rmsdlength}
\end{figure}

To visualize these results, the global RMSD of the top decoy is plotted against loop length for each method in Figure \ref{fig:rmsdlength}. It is clear that the prediction difficulty and the variance of prediction RMSDs tend to increase with loop length, with methods consistently achieving $<$2 \AA~RMSD accuracy only for the shortest loops ($\le 6$ residues). This is sensible since the size of the conformational space increases with loop length, with long loops ($>12$ residues) often posing a challenge for methods to sample adequately \citep{li2011vsgb}. The plots also indicate that hardest targets for a given loop length tend to be those from multiple conformations, especially for the two most accurate methods (NGK and PETALS). The average lengths of loop targets in the `Single conf.' and  `Multiple conf.' categories are similar (9.7 vs.~10.0 residues). The detailed results for each target individually are given in Table S1 of the Supporting Information.

If one is allowed to select the best prediction among \textit{all} targets for a loop instance, then results for loops with multiple conformations improve dramatically (e.g., taking the lowest RMSD of all decoys generated from 6xluB, 7kdkC, and 7kdlA together as the result for the loop 130--140); the average global RMSD for the top decoy in multiple conformation loops decreases to just 1.05 \AA~for NGK and 1.74 \AA~for PETALS. However, this is generally not a realistic scenario in practice, as often just a single template would be available for constructing predictions. In this sense, our findings on the difficulty of predicting multiple conformation loops are less categorical compared to \citet{marks2018predicting} for the targets in this S protein dataset. For these S protein targets, multiple conformation loops are more difficult to predict when a single template is used, but not when we can choose the best prediction among all available templates; for the dataset considered by \citet{marks2018predicting}, the difficulty still remained when choosing the best prediction among all available templates, albeit accounting for less possible within-cluster variation as their clusters had much less representation in the PDB. 

\newpage
In addition to loop length, we also examine whether the cluster characteristics, namely their size (as measured by the number of chains) and breadth (as measured by the average within-cluster RMSD in Figure \ref{fig:withinrmsd}), are associated with prediction difficulty. For each method, we consider a target to be successfully predicted if the top decoy has a global RMSD of $<$2~\AA, and to be a failure otherwise. Based on this criterion, DiSGro, NGK, PETALS, and Sphinx had 25 (48\%), 35 (67\%), 31 (60\%), and 25 (48\%) successes, respectively, out of the 52 loop targets representing  conformational clusters defined by at least four chains and two distinct PDB codes. We use the Welch $t$-test to provide a simple assessment of whether the mean of each variable is significantly different between successes and failures, and the results are shown in Table \ref{tab:ttest} for the four methods. The sign of the $t$-statistic indicates whether successes (positive $t$-statistic) or failures (negative $t$-statistic) are associated with larger values of that variable; e.g., the $t$-statistics for loop length are all negative, so successes are associated with shorter loop lengths as expected from Figure \ref{fig:rmsdlength}. Each of the three variables is significantly associated with prediction success ($p < 0.01$ for all tests, except cluster size for the Sphinx method with $p=0.011$). Targets with longer loop lengths, smaller cluster sizes, and larger cluster breadths tend to be more difficult to predict successfully, regardless of which loop modeling method is used.

\begin{table}[!htbp] 
	\caption{Comparing prediction successes and failures of the four methods, according to loop length, cluster size, and cluster breadth. Prediction success is defined as a global RMSD of $<$2~\AA~for the top decoy. The Welch $t$-statistics (with degrees of freedom in brackets) and $p$-values for each variable are shown. Positive $t$-statistics indicate that successes have a larger mean than failures. 	The tests are based on the loop targets representing  conformational clusters defined by at least four chains and two distinct PDB codes.	\strut} 
	\label{tab:ttest} 
	\centering
	\begin{tabular}{@{\extracolsep{5pt}}lcccc} 
		 \textbf{Variables} & \multicolumn{4}{c}{\textbf{Welch $t$-test results}} \\ 
		\cline{2-5} \\
		[-1.8ex] 
		 & DiSGro & NGK & PETALS & Sphinx \\ 
		\hline \\[-1.8ex] 
		Loop length & $t(41.1) = -6.32$ & $t(30.3) = -3.53$ & $t(36.5) = -5.20$ & $t(49.4) = -3.23$ \\ 
		& $p < 0.001$ & $p = 0.0015$ & $p < 0.001$ & $p = 0.0022$ \\ 
		& & & & \\ 
		Cluster size & $t(49.2) = 4.18$ & $t(31.1) = 3.91$ & $t(41.9) = 3.10$ & $t(50.0) = 2.63$ \\ 
		& $p < 0.001$ & $p < 0.001$ & $p = 0.0034$ & $p = 0.011$ \\ 
		& & & & \\ 
		Cluster breadth & $t(47.7) = -4.62$ & $t(23.1) = -3.52$ & $t(35.0) = -3.08$ & $t(48.2) = -2.94$ \\ 
		& $p < 0.001$ &  $p = 0.0018$ & $ p = 0.0040$ & $p = 0.0050$ \\ 
		\hline \\[-1.8ex] 
	\end{tabular} 
\end{table} 

Next we focus on the loop instances with multiple distinct conformations, to assess how well the decoys generated from a specific PDB input can represent \textit{all} the known conformations for that loop instance. Taking the loop 130--140 for example: the decoys generated using 6xluB are compared to the loop structures in the clusters represented by 6xluB, 7kdkC, 7kdlA, and the RMSD to the closest structure in each cluster is recorded; the average of the RMSDs to these three clusters then provides an overall result for 6xluB; the same is done using the decoys from 7kdkC and 7dklA. The results are summarized in Table \ref{tab:rmsdmulti} using the same RMSD metrics, averaged over the targets in the multiple conformation category. This task is noticeably more challenging than the prior prediction task, as evidenced by RMSDs in Table \ref{tab:rmsdmulti} which are all larger than the corresponding values in the `Multiple conf.' rows of Table \ref{tab:rmsdpred} for all four methods. While the top decoy RMSDs are expected to be increase relative to Table \ref{tab:rmsdpred}, a substantial increase still occurs when taking the entire decoy set (`Min.' column, e.g., 1.07 to 2.18 \AA~global RMSD for NGK) and when allowing methods to choose the top five decoys (`Top-5' column, e.g., 1.59 to 2.85 \AA~global RMSD for NGK), whether considering local or global RMSD. This suggests that building the loop using the atomic environment of a single structural template may preclude the methods from being able to locate and predict all the possible loop conformations; \citet{marks2018predicting} observed a similar phenomenon in their dataset. The detailed results for each loop target individually are given in Table S2 of the Supporting Information.

\begin{table}[ht]
		\caption{RMSD metrics for the loop instances with multiple conformations. 
		The loop backbone RMSDs shown are averaged over the targets in the multiple conformation category, where decoys generated from each target are compared to all known conformations for that loop instance and RMSDs are calculated to the closest structure in each cluster. 
		The columns `Min.', `Top', and `Top-5' refer respectively to the lowest RMSD among the 500 decoys, RMSD of the top-ranked decoy, and lowest RMSD among the top-five ranked decoys.}	
	\centering
	\begin{tabular}{lcccccc}
		& \multicolumn{3}{c}{Local RMSD} & \multicolumn{3}{c}{Global RMSD}   \\ \cmidrule(lr){2-4} \cmidrule(lr){5-7}
		\textbf{Method} & \textbf{Min.} & \textbf{Top} & \textbf{Top-5} & \textbf{Min.} & \textbf{Top} & \textbf{Top-5} \\ 
		\hline
		DiSGro & 1.36 & 2.40 & 2.00 & 2.50 & 4.76 & 4.05 \\ 
		NGK & 1.19 & 2.01 & 1.65 & 2.18 & 3.84 & 2.85 \\ 
		PETALS & 1.28 & 2.11 & 1.86 & 2.56 & 4.26 & 3.60 \\ 
		Sphinx & 1.14 & 2.24 & 1.80 & 2.28 & 4.70 & 3.65 \\ 
		\hline
	\end{tabular}
	\label{tab:rmsdmulti}
\end{table}

The multiple conformation loop instances in the RBD were not more difficult to predict. Methods located known conformations from their loop targets at a comparable level of accuracy versus those outside the RBD; e.g., average global RMSDs for assessing the representation of all the conformations in the top five decoys were 2.55 vs.~3.07 \AA~for NGK and 4.21 vs.~3.94 for DiSGro. The average length of these loop targets in the RBD is 9.9 residues, and similar to the average length (10.0) among all multiple conformation targets. The loop regions with sequence variants in the PDB had little structural variability (Table \ref{tab:mut}) and were not expected to pose additional challenges for the loop modeling methods. Detailed results for each sequence variant confirm this, and are provided in Table S3 of the Supporting Information.

Five loop targets were omitted from the above analyses due to challenges encountered when running the methods. The two very long loops in the set, namely 146--168 and 783--816, were particularly difficult, with DiSGro and Sphinx unable to generate decoys possibly due to their lengths.  The 146--168 loop has two conformations, both of which could be predicted moderately well by PETALS (top decoy global RMSDs: 2.18 for 6zgiB conformation, 2.39 for 7dddC conformation) and NGK (top decoy global RMSDs: 2.80 for 6zgiB conformation, 2.45 for 7dddC conformation).  The length 34 loop (783--816) is very challenging, and no method could give useful results (top decoy global RMSDs: 26.8 for NGK, 12.0 for PETALS).  The Sphinx webserver was also unable to generate decoys for 31--46 and 320--324 (6xm0A conformation) possibly due to a lack of suitable templates.  Further, some of Sphinx's jobs were unable to complete the full SOAP-Loop ranking steps; thus, we used the 500 SOAP-Loop ranked decoys if they were available, and otherwise selected its top 500 decoys from the coarse-grained ranking stage for our analysis.  Detailed results for these five targets are provided in Table S4 of the Supporting Information.

\section{Conclusion}

In this paper, we studied the conformations of loops in the SARS-CoV-2 S protein.  We extracted all SARS-CoV-2 S protein loop regions, examined their sequence and structural variability based on the available structures in the PDB, and applied loop modeling methods to assess how well the loop conformations could be predicted.  Forty-four loop regions were identified, and as the structure of the S protein has been experimentally solved many times, 17 loop instances were observed to have substantive structural variability and be able to adopt multiple distinct conformations according to a cluster analysis. The clusters gave insights into the amount of structural uncertainty present in these loops, and there were quantifiable differences in their sizes and breadths.

Loops' frequent association with protein function, together with their more disordered nature compared to regular secondary structures, means that their accurate modeling is an important problem in structural biology. Specifically for the S protein, loop regions we identified include 475--487 and 495--506, which correspond to key loops known to be involved in binding with ACE2. These are referred to as `Loop 3' and `Loop 4' in \citet{williams2021molecular}, where molecular dynamics simulations revealed `Loop 3' to be highly flexible in the unbound state, including the possibility of a conformation that inhibits ACE2 binding. Interestingly, our results also showed that 475--487 was one of the most difficult loops to predict, with all four methods struggling with the 6xm0B template (global RMSD of top decoy $>10$ \AA, Table S2). Exploring the conformational variability of `Loop 3' thus provides a fuller range of structural states that the development of therapeutics might target before the S protein binds to ACE2 \citep{williams2021molecular}. More generally, high-quality loop models are a crucial part of protein structures used in the computational drug discovery process \citep{muhammed2019homology}.

We found that the structurally flexible loops with multiple conformations in the S protein tended to be more challenging for loop modeling methods to predict a correct structure, compared to relatively inflexible loops with a single conformation. Prediction accuracies were strongly associated with loop length, due to the larger conformational space of  longer loops. Further, it was very challenging for methods to predict all known conformations from a single structural template. Our results thus highlight limitations of current loop prediction methods, most of which were designed to predict a single `correct' conformation. These echo some of the findings in \citet{marks2018predicting}, but with some important distinctions. First, we were able to more fully consider cluster size and breadth in the analysis, thanks to the large number of S protein chains in the PDB. Second, we did not construct a curated set of high and low flexibility loops specifically, but rather considered all S protein loops which cover a wider range of loop structural variability.  In effect, a much larger proportion of loops (17 of 44 in our study) may be considered highly flexible, if other structures were to be solved this many times. Third, the multiple conformation targets in our dataset were easier to predict than those of \citet{marks2018predicting} when allowing the best decoys across all structural templates to be chosen. Overall, this work provides insight into the abilities of current loop prediction methods for a key protein associated with the ongoing COVID-19 disease, and identifies the loops where structural flexibility could play a role as the SARS-CoV-2 virus continues to evolve. Future study in loop modeling protocols might better incorporate multiple conformation loops in their training data and improve prediction accuracies for longer loops.

Finally, we note one limitation of this study, namely our focus on loops rather than more global protein structure.  In this sense, more global structural variability across S protein chains may have hindered the ability of methods to locate all the distinct loop conformations from a single input structure, since the rest of the protein chain is held fixed. Additionally, we found the observable changes to loop structures from known sequence variants in the PDB to be small. There could be more global structural changes due to mutation not detected by the current analysis, for example the D614G mutation \citep{yurkovetskiy2020structural}.  Nonetheless, loops deserve careful study in their own right, due to their functional importance.  Further study could focus on larger-scale variability in the S protein structure, leveraging the rich source of experimental data available in the PDB to better understand COVID-19.

\section*{Acknowledgements}
This work was partially supported by Discovery Grant RGPIN-2019-04771 from the Natural Sciences and Engineering Research Council of Canada.

\section*{Data availability statement}
The data that support the findings of this study are openly available in the RCSB Protein Data Bank.

\appendix 

\section{B-factor analysis}

Let 
$(x_{11}, y_{11}, z_{11}), \ldots, (x_{1N}, y_{1N}, z_{1N})$
and 
$(x_{21}, y_{21}, z_{21}), \ldots, (x_{2N}, y_{2N}, z_{2N})$ denote the measured backbone coordinates for the pair being compared, with corresponding B-factors denoted by $B_{11}, \ldots B_{1N}$ and $B_{21}, \ldots B_{2N}$. 

Since the B-factor is defined as $B = 8 \pi^2 \left<u^2\right>$, a Gaussian approximation gives the variance in each measured $x$, $y$, and $z$ coordinate as $B/(3\cdot 8 \pi^2)$.  For the $i$-th atom, the coordinate difference between the pair is a random vector $(H_{xi}, H_{yi}, H_{zi})$ with a multivariate Gaussian distribution with mean vector $(x_{1i} - x_{2i}, y_{1i} - y_{2i}, z_{1i} - z_{2i})$ and a diagonal covariance matrix with the value $\sigma^2_i = (B_{1i} + B_{2i} ) /(3\cdot 8 \pi^2)$ along its diagonal.

By the properties of the multivariate Gaussian, 
$$
\frac{(H_{xi} - (x_{1i} - x_{2i}))^2 +  (H_{yi} - (y_{1i} - y_{2i}))^2  + (H_{zi} - (z_{1i} - z_{2i}))^2}{\sigma^2_i}
$$
has a chi-squared distribution with 3 degrees of freedom, denoted $\chi^2_3$. Similarly, considering all the atoms together, a 
$\chi^2_{3N}$ random variable is defined by
	$$
	\sum_{i=1}^N  \frac{(H_{xi} - (x_{1i} - x_{2i}))^2 +  (H_{yi} - (y_{1i} - y_{2i}))^2  + (H_{zi} - (z_{1i} - z_{2i}))^2}{\sigma^2_i}.
	$$
The pair of loop backbones are not different if it is plausible that $(H_{xi}, H_{yi}, H_{zi}) = (0,0,0)$ for all $N$ atoms, i.e., all the coordinate differences are zero. This corresponds to computing the statistic 
$$
T = \sum_{i=1}^N  \frac{(x_{1i} - x_{2i})^2 +  (y_{1i} - y_{2i})^2  + (z_{1i} - z_{2i})^2}{\sigma^2_i}
$$
and comparing $T$ to the quantiles of the chi-squared distribution with $3N$ degrees of freedom. Taking a significance level of $\alpha = 0.05$, let $c$ denote the 0.95 quantile of the $\chi^2_{3N}$ distribution. Then the pair is considered significantly different if $T > c$.

\section{Updated scoring function for PETALS algorithm}

In this work we also tested a strategy for improving the energy function accuracy of the PETALS algorithm, in its ability to rank generated loop decoys. The set of structures used for training is the same as that described in \citet{wong2017fast}, namely, the CulledPDB list by PISCES \citep{wang2003pisces} on March 14, 2015 with maximum 20\% sequence identity, resolution 2.0 \AA, and $R$-factor cutoff 0.25, thus ensuring no SARS-CoV-2 S protein structures were present.  Loop regions were extracted via DSSP, from which we compiled 10786 loops with lengths ranging from 5 to 10 residues.

The PETALS algorithm was first used to generate 200 decoys for each loop, and for each decoy we computed: RMSD to the native conformation, 210 distance-based energy terms corresponding to each pair of atom types defined in DiSGro's energy function \citep{tang2014fast}, and a backbone torsion term \citep{wong2017fast}. 
We then define $\hat{y}_{ij}$ as the predicted energy of the $i$-th loop's $j$-th decoy according to 
\begin{equation} \label{eq:1}
\hat{y}_{ij} =  T_{ij} + \sum_{k=1}^{210} \beta_{k} E_{ijk}, \nonumber 
\end{equation}
where $\beta_{k}$'s are coefficients associated with each energy term $E_{ijk}$ to be trained, and $T_{ij}$ is the torsion term. 
Then define the square-error loss function
\begin{equation} \label{eq:2}
\sum_{i=1}^{N} \sum_{j=1}^{200} w_{ij} \left( f\left(\hat{y}_{ij} \right) - f(\text{RMSD}_{ij}) \right)^{2},
\end{equation}
where 
$\text{RMSD}_{ij}$ is the RMSD to native and $w_{ij}$ is the weight associated with the $i$-th loop's $j$-th decoy, $N$ is the number of training loops, and $f$ is a mapping function 
associated with the rank of that decoy.
The decoys with the lowest RMSDs are the ones that 
best resemble the true conformation; thus the goal 
is to 
train the $\beta_k$'s to minimize this loss function so that the rankings of the predicted energies and the rankings of the RMSD values match as closely as possible.

We chose $f(\cdot)$ to be a function that  
maps values into quantile bins.  Specifically, we ranked the 200 predicted energies
$\left\{ \hat{y}_{ij} \right\}_{j=1}^{200}$ from smallest to largest, then assigning $f=1$ to the best 10\%, $f=2$ to the next 10\%, until $f=10$ for the last 10\%.  We  ranked the 200 RMSD values $\{ RMSD_{ij}\}_{j=1}^{200}$ and assigned values of $f$ the same way.
Positive weights $w_{ij}$ were assigned to the top five quantile bins, with higher weights for the better ranked predicted energies: $1.0$ for the best 10\%, $0.9$ for the next 10\%, until $0.6$ for 5th quantile bin, and zero for the rest.  We used 80\% of the loops as training data and 20\% as validation data. As gradient information was unavailable due to the discrete nature of the model, the PySwarms \citep{pyswarmsJOSS2018} implementation of Particle Swarm Optimization was used to minimize the square error loss function in Equation (\ref{eq:2}).

\clearpage
\renewcommand{\thetable}{S\arabic{table}}	
\setcounter{table}{0}

%	\title{Supporting Information for ``Conformational variability of loops in the SARS-CoV-2 spike protein'' by Samuel W.K. Wong and Zongjun Liu}

\subsection*{Supporting Information for ``Conformational variability of loops in the SARS-CoV-2 spike protein''}	
%	\date{}
%	\maketitle
	
%	\noindent This supporting information file contains the following:
	\begin{itemize}
		\item Table S1: Detailed loop prediction accuracies of the four methods. %, for each of the multiple conformation targets.
		\item Table S2: Detailed RMSD metrics for the loop instances with multiple conformations, comparing the decoys generated from a given target to all known conformations for that loop instance.
		%	\item Table S3: Detailed loop reconstruction accuracies of the four methods.
		\item Table S3: Detailed RMSD metrics for the loop regions with sequence variants.
		\item Table S4: Detailed RMSD metrics for the five loop targets omitted from the main analyses.
	\end{itemize}
	
	\newpage
	
	\newgeometry{left=0.5in,right=0.5in,top=1in, bottom=1in}
	% latex table generated in R 3.6.3 by xtable 1.8-4 package
	% Fri Apr 30 00:19:03 2021
		\footnotesize	
	\begin{longtable}[ht]{ll|rrr|rrr|rrr|rrr}
		%\begin{table}[ht]
		\caption{RMSD metrics for assessing the prediction accuracy of the four methods.
			The loop backbone RMSDs are shown for each of the 66 targets. %37 multiple conformation targets. 
			The columns `Min.', `Top1', and `Top5' refer respectively to the lowest RMSD among the 500 decoys, RMSD of the top-ranked decoy, and lowest RMSD among the top-five ranked decoys, where each is calculated to the closest structure among all chains containing that loop instance.  The PDB column indicates the representative chain used to generate loop decoys. %, while `Comp.' indicates the representative chain containing the loop conformation to which the decoys are being compared. 
			For example, 130--140 has three distinct loop conformations, represented by the PDB chains 6xluB, 7kdkC, and 7kdlA; using 6xluB as input, the top decoy of the DiSGro method had local RMSD 1.02 \AA~to the closest structure among all chains (i.e., in all three clusters) that contained the 130--140 loop instance.} \\
		%For example, 130--140 has three distinct loop conformations, represented by the PDB chains 6xluA, 7kdkC, and 7kdlC; using 6xluA as input, the top decoy of the DiSGro method was able to reconstruct the 130--140 loop in 6xluA with RMSD 1.75 \AA.} \\
		%	\centering
		%	\begin{tabular}{ll|rrr|rrr|rrr|rrr}
		\multicolumn{13}{c}{\textbf{Local RMSD}} \\
		\hline
		& & \multicolumn{3}{c|}{DiSGro} & \multicolumn{3}{c|}{NGK} &
		\multicolumn{3}{c|}{PETALS} & \multicolumn{3}{c}{Sphinx}   \\
		\hline
		Region & PDB & Min. & Top1 & Top5 & Min. & Top1 & Top5 & Min. & Top1 & Top5 & Min. & Top1 & Top5 \\ 
		\hline
		56-60 & 6xr8A & 0.10 & 0.21 & 0.19 & 0.05 & 0.11 & 0.09 & 0.08 & 0.17 & 0.17 & 0.08 & 0.29 & 0.18 \\ 
		108-116 & 6zoxB & 1.05 & 1.85 & 1.56 & 0.22 & 0.90 & 0.73 & 0.36 & 0.93 & 0.87 & 0.80 & 2.38 & 1.45 \\ 
		130-140 & 6xluB & 0.57 & 1.02 & 1.01 & 0.16 & 0.67 & 0.32 & 0.35 & 0.67 & 0.47 & 0.77 & 2.22 & 1.19 \\ 
		130-140 & 7kdkC & 1.35 & 2.50 & 2.37 & 0.68 & 1.96 & 1.19 & 1.32 & 2.12 & 1.86 & 1.21 & 2.09 & 1.98 \\ 
		130-140 & 7kdlA & 1.43 & 2.58 & 1.69 & 0.72 & 3.14 & 1.52 & 0.61 & 2.28 & 1.48 & 1.05 & 2.96 & 1.64 \\ 
		172-187 & 6zp0B & 1.87 & 5.28 & 3.47 & 2.19 & 3.76 & 2.31 & 1.77 & 4.58 & 4.58 & 1.51 & 2.89 & 2.09 \\ 
		172-187 & 7df3B & 1.86 & 4.99 & 3.72 & 0.41 & 0.53 & 0.41 & 1.16 & 1.67 & 1.63 & 1.67 & 2.42 & 2.42 \\ 
		210-222 & 6vxxA & 1.47 & 3.88 & 1.47 & 0.84 & 2.91 & 1.99 & 0.93 & 2.79 & 1.07 & 1.07 & 4.05 & 4.05 \\ 
		230-236 & 6vxxA & 0.27 & 0.89 & 0.76 & 0.12 & 0.17 & 0.17 & 0.18 & 0.28 & 0.28 & 0.27 & 0.98 & 0.45 \\ 
		245-263 & 6zgiB & 2.07 & 4.76 & 2.91 & 1.54 & 3.55 & 3.55 & 3.11 & 4.67 & 4.42 & 2.50 & 4.07 & 4.07 \\ 
		280-284 & 6x79B & 0.08 & 0.13 & 0.13 & 0.05 & 0.09 & 0.07 & 0.07 & 0.14 & 0.12 & 0.07 & 0.15 & 0.11 \\ 
		304-310 & 7a4nB & 0.21 & 0.87 & 0.37 & 0.09 & 0.16 & 0.10 & 0.16 & 0.26 & 0.16 & 0.27 & 0.33 & 0.33 \\ 
		320-324 & 6zoxC & 0.11 & 0.44 & 0.39 & 0.06 & 0.14 & 0.12 & 0.08 & 0.14 & 0.14 & 0.09 & 0.14 & 0.10 \\ 
		329-338 & 6x29A & 0.45 & 1.45 & 1.08 & 0.16 & 0.20 & 0.17 & 0.39 & 1.41 & 0.72 & 0.52 & 1.96 & 1.35 \\ 
		329-338 & 7kdlB & 0.52 & 1.28 & 0.71 & 0.35 & 1.32 & 1.13 & 0.53 & 1.33 & 0.96 & 0.65 & 2.16 & 1.95 \\ 
		343-348 & 6zgeC & 0.25 & 1.67 & 1.37 & 0.07 & 0.12 & 0.10 & 0.17 & 0.23 & 0.18 & 0.19 & 1.41 & 0.23 \\ 
		370-375 & 6vxxA & 0.19 & 0.30 & 0.30 & 0.08 & 0.59 & 0.16 & 0.12 & 0.26 & 0.16 & 0.16 & 1.75 & 1.21 \\ 
		370-375 & 6zgiC & 0.42 & 1.03 & 1.03 & 0.35 & 0.88 & 0.78 & 0.32 & 0.63 & 0.42 & 0.43 & 1.54 & 1.51 \\ 
		380-394 & 6x79B & 1.45 & 3.17 & 1.45 & 0.53 & 0.58 & 0.55 & 0.64 & 0.72 & 0.72 & 0.63 & 0.70 & 0.70 \\ 
		380-394 & 7kdlC & 2.06 & 2.74 & 2.74 & 0.41 & 0.49 & 0.43 & 0.57 & 0.90 & 0.57 & 0.42 & 0.55 & 0.52 \\ 
		410-416 & 6zoxB & 0.71 & 1.88 & 1.76 & 0.18 & 0.20 & 0.18 & 0.20 & 0.32 & 0.20 & 0.35 & 0.62 & 0.62 \\ 
		410-416 & 7kdkA & 0.28 & 1.49 & 1.46 & 0.17 & 0.22 & 0.17 & 0.15 & 0.25 & 0.25 & 0.08 & 0.87 & 0.19 \\ 
		422-430 & 6xm0B & 0.61 & 1.57 & 1.02 & 0.74 & 1.53 & 1.32 & 0.87 & 1.40 & 1.32 & 0.64 & 1.07 & 1.01 \\ 
		422-430 & 6xr8B & 0.55 & 0.84 & 0.84 & 0.17 & 0.22 & 0.22 & 0.28 & 0.74 & 0.38 & 0.24 & 0.29 & 0.25 \\ 
		438-451 & 6xr8A & 1.53 & 2.41 & 2.41 & 2.05 & 3.03 & 3.03 & 2.40 & 2.84 & 2.84 & 0.32 & 0.47 & 0.39 \\ 
		438-451 & 7kdlB & 1.99 & 3.29 & 2.77 & 1.99 & 3.17 & 2.82 & 2.10 & 3.58 & 2.54 & 1.04 & 4.45 & 2.56 \\ 
		454-472 & 6zgeC & 1.42 & 3.52 & 3.00 & 1.89 & 5.67 & 4.35 & 2.31 & 3.35 & 3.35 & 0.69 & 3.24 & 3.24 \\ 
		475-487 & 6xm0B & 1.65 & 3.18 & 2.97 & 1.37 & 3.06 & 1.89 & 1.43 & 2.61 & 2.13 & 1.47 & 3.90 & 2.78 \\ 
		475-487 & 7dddA & 1.03 & 2.67 & 2.24 & 0.52 & 0.75 & 0.75 & 1.69 & 2.01 & 2.01 & 1.21 & 2.74 & 2.05 \\ 
		495-506 & 6xm0B & 1.89 & 2.36 & 2.36 & 1.26 & 2.04 & 1.64 & 1.81 & 2.09 & 2.09 & 0.73 & 2.82 & 2.06 \\ 
		495-506 & 6zp0A & 1.79 & 2.39 & 2.12 & 0.41 & 3.03 & 2.64 & 0.95 & 2.38 & 2.38 & 0.39 & 0.52 & 0.44 \\ 
		495-506 & 7kdlB & 1.87 & 3.06 & 2.81 & 1.31 & 2.97 & 1.91 & 1.49 & 2.73 & 2.65 & 0.48 & 2.97 & 1.06 \\ 
		517-523 & 6xm0A & 0.61 & 0.96 & 0.93 & 0.37 & 0.46 & 0.45 & 0.47 & 1.17 & 1.04 & 0.27 & 0.95 & 0.95 \\ 
		517-523 & 6xm0B & 0.96 & 1.30 & 1.26 & 0.99 & 1.06 & 1.06 & 0.94 & 1.29 & 1.29 & 0.89 & 1.29 & 1.19 \\ 
		517-523 & 6xm3A & 0.35 & 1.25 & 0.35 & 0.24 & 0.34 & 0.33 & 0.24 & 1.12 & 1.07 & 0.29 & 0.34 & 0.34 \\ 
		517-523 & 6zoxA & 0.44 & 1.10 & 1.00 & 0.19 & 0.25 & 0.22 & 0.20 & 0.44 & 0.28 & 0.25 & 1.23 & 0.58 \\ 
		526-537 & 6x29B & 1.47 & 2.21 & 1.89 & 0.27 & 0.47 & 0.32 & 0.41 & 0.41 & 0.41 & 0.65 & 0.75 & 0.75 \\ 
		526-537 & 7ad1B & 1.21 & 1.59 & 1.54 & 0.32 & 0.92 & 0.50 & 0.41 & 0.41 & 0.41 & 0.71 & 1.52 & 0.84 \\ 
		555-564 & 7kdkC & 0.91 & 2.01 & 1.32 & 0.24 & 0.30 & 0.30 & 0.34 & 1.49 & 0.54 & 0.42 & 0.51 & 0.51 \\ 
		578-583 & 6zoxB & 0.35 & 0.67 & 0.42 & 0.10 & 0.45 & 0.32 & 0.19 & 0.31 & 0.27 & 0.11 & 0.28 & 0.17 \\ 
		600-608 & 7kdlA & 0.33 & 0.96 & 0.96 & 0.12 & 1.37 & 0.17 & 0.35 & 1.82 & 0.35 & 0.63 & 2.09 & 0.72 \\ 
		614-620 & 6xm4C & 0.43 & 2.17 & 1.41 & 0.42 & 1.71 & 1.68 & 0.36 & 0.86 & 0.86 & 0.33 & 1.62 & 0.40 \\ 
		624-641 & 6xm0B & 2.09 & 5.15 & 2.88 & 0.69 & 0.69 & 0.69 & 2.63 & 4.71 & 3.79 & 2.89 & 4.16 & 4.16 \\ 
		656-663 & 7kdkB & 0.25 & 0.58 & 0.58 & 0.17 & 0.21 & 0.21 & 0.24 & 0.34 & 0.32 & 0.37 & 0.83 & 0.67 \\ 
		697-710 & 6vxxB & 1.22 & 1.93 & 1.85 & 1.07 & 2.38 & 1.46 & 1.47 & 2.54 & 1.93 & 1.16 & 2.26 & 1.84 \\ 
		825-836 & 6xluB & 1.38 & 2.64 & 1.77 & 1.44 & 2.30 & 2.04 & 1.64 & 2.62 & 2.51 & 1.34 & 2.78 & 2.24 \\ 
		825-836 & 6xm3B & 1.18 & 3.42 & 2.92 & 0.51 & 1.91 & 0.55 & 0.70 & 1.15 & 1.15 & 1.23 & 1.71 & 1.71 \\ 
		825-836 & 6xm3C & 1.41 & 2.70 & 1.47 & 0.62 & 1.91 & 1.34 & 0.75 & 1.92 & 1.42 & 1.72 & 3.10 & 2.79 \\ 
		825-836 & 6zgiA & 1.15 & 1.75 & 1.75 & 1.28 & 2.10 & 2.06 & 1.61 & 1.97 & 1.97 & 1.14 & 2.77 & 1.49 \\ 
		841-848 & 6xluC & 0.54 & 1.89 & 0.99 & 0.37 & 2.05 & 1.81 & 0.68 & 1.43 & 0.71 & 0.42 & 0.98 & 0.45 \\ 
		841-848 & 6xm3B & 0.83 & 1.98 & 1.98 & 0.77 & 2.35 & 1.31 & 0.75 & 1.51 & 1.38 & 0.56 & 2.21 & 0.94 \\ 
		841-848 & 6xm4B & 0.68 & 1.99 & 1.49 & 0.56 & 1.62 & 1.24 & 1.00 & 2.24 & 1.79 & 0.62 & 1.40 & 0.94 \\ 
		841-848 & 6zgeB & 0.45 & 2.02 & 0.45 & 0.48 & 0.77 & 0.77 & 0.70 & 2.02 & 1.41 & 0.36 & 0.52 & 0.43 \\ 
		841-848 & 7dddB & 0.92 & 1.81 & 1.77 & 0.29 & 1.95 & 1.86 & 0.90 & 1.63 & 1.10 & 0.47 & 2.19 & 1.09 \\ 
		862-866 & 6zoxB & 0.08 & 1.37 & 0.63 & 0.05 & 0.10 & 0.08 & 0.06 & 0.12 & 0.06 & 0.07 & 0.22 & 0.15 \\ 
		891-897 & 7a4nB & 0.16 & 0.47 & 0.20 & 0.29 & 0.53 & 0.52 & 0.26 & 0.55 & 0.43 & 0.23 & 1.02 & 0.46 \\ 
		891-897 & 7kdkB & 0.27 & 0.29 & 0.27 & 0.42 & 0.83 & 0.82 & 0.22 & 0.33 & 0.33 & 0.32 & 0.97 & 0.76 \\ 
		908-913 & 7a4nB & 0.18 & 0.45 & 0.25 & 0.08 & 0.13 & 0.11 & 0.13 & 0.39 & 0.32 & 0.12 & 0.23 & 0.19 \\ 
		968-976 & 6xraC & 0.25 & 0.44 & 0.33 & 0.20 & 0.20 & 0.20 & 0.34 & 0.52 & 0.42 & 0.63 & 1.45 & 1.26 \\ 
		968-976 & 6zp0C & 0.56 & 1.11 & 1.11 & 0.13 & 0.19 & 0.16 & 0.26 & 1.41 & 0.29 & 0.90 & 1.92 & 1.85 \\ 
		1033-1046 & 7kdkA & 1.64 & 3.03 & 2.36 & 0.92 & 3.45 & 3.12 & 2.22 & 2.91 & 2.91 & 2.27 & 3.95 & 3.35 \\ 
		1106-1112 & 7kdkC & 0.32 & 1.02 & 0.99 & 0.17 & 0.21 & 0.17 & 0.28 & 0.99 & 0.90 & 0.28 & 1.08 & 0.50 \\ 
		1124-1132 & 6xm0A & 0.67 & 1.13 & 0.97 & 0.30 & 0.34 & 0.34 & 0.38 & 0.68 & 0.50 & 0.50 & 1.77 & 1.77 \\ 
		1124-1132 & 6xraC & 0.96 & 2.06 & 1.66 & 1.18 & 1.37 & 1.33 & 0.98 & 1.83 & 1.82 & 1.26 & 1.35 & 1.35 \\ 
		1135-1141 & 6xraC & 0.55 & 1.53 & 1.19 & 0.74 & 0.94 & 0.91 & 0.60 & 1.51 & 1.32 & 0.75 & 0.85 & 0.85 \\ 
		1135-1141 & 7kdkB & 0.21 & 0.58 & 0.35 & 0.11 & 0.16 & 0.16 & 0.21 & 0.25 & 0.25 & 0.19 & 0.47 & 0.46 \\ 
		\hline
		\multicolumn{13}{c}{~~~} \\
		\multicolumn{13}{c}{\textbf{Global RMSD}} \\
		\hline
		Region & PDB & Min. & Top1 & Top5 & Min. & Top1 & Top5 & Min. & Top1 & Top5 & Min. & Top1 & Top5 \\ 
		\hline		
		56-60 & 6xr8A & 0.20 & 0.29 & 0.29 & 0.11 & 0.20 & 0.13 & 0.20 & 0.33 & 0.27 & 0.26 & 0.46 & 0.36 \\ 
		108-116 & 6zoxB & 1.15 & 2.54 & 2.07 & 0.71 & 1.16 & 1.01 & 0.50 & 1.08 & 0.99 & 1.41 & 5.25 & 1.61 \\ 
		130-140 & 6xluB & 1.08 & 1.30 & 1.23 & 0.41 & 0.87 & 0.60 & 0.63 & 0.92 & 0.64 & 1.03 & 2.54 & 1.64 \\ 
		130-140 & 7kdkC & 1.75 & 3.33 & 3.08 & 0.85 & 2.13 & 1.34 & 1.80 & 3.47 & 2.19 & 1.82 & 3.11 & 3.00 \\ 
		130-140 & 7kdlA & 1.63 & 3.18 & 2.10 & 1.41 & 4.76 & 2.19 & 0.83 & 5.09 & 2.49 & 1.36 & 3.98 & 1.96 \\ 
		172-187 & 6zp0B & 3.92 & 10.71 & 5.85 & 2.69 & 10.53 & 2.78 & 2.17 & 10.14 & 9.88 & 3.06 & 10.90 & 6.36 \\ 
		172-187 & 7df3B & 3.12 & 9.69 & 7.78 & 0.70 & 0.80 & 0.70 & 1.36 & 2.58 & 2.20 & 3.82 & 8.48 & 7.38 \\ 
		210-222 & 6vxxA & 1.81 & 6.99 & 1.82 & 1.38 & 5.61 & 2.89 & 1.09 & 3.75 & 1.28 & 2.12 & 6.77 & 6.77 \\ 
		230-236 & 6vxxA & 0.39 & 1.04 & 0.89 & 0.16 & 0.21 & 0.21 & 0.31 & 0.35 & 0.35 & 0.43 & 1.32 & 0.69 \\ 
		245-263 & 6zgiB & 2.66 & 11.71 & 5.61 & 1.58 & 11.49 & 11.02 & 5.58 & 9.33 & 9.15 & 6.91 & 13.81 & 12.74 \\ 
		280-284 & 6x79B & 0.16 & 0.21 & 0.21 & 0.08 & 0.12 & 0.12 & 0.14 & 0.22 & 0.22 & 0.15 & 0.28 & 0.21 \\ 
		304-310 & 7a4nB & 0.25 & 0.90 & 0.42 & 0.15 & 0.21 & 0.15 & 0.25 & 0.30 & 0.28 & 0.42 & 0.54 & 0.42 \\ 
		320-324 & 6zoxC & 0.15 & 0.53 & 0.53 & 0.17 & 0.29 & 0.24 & 0.14 & 0.20 & 0.20 & 0.36 & 0.57 & 0.50 \\ 
		329-338 & 6x29A & 0.58 & 1.50 & 1.11 & 0.36 & 0.38 & 0.38 & 0.62 & 2.05 & 0.85 & 1.04 & 3.14 & 1.89 \\ 
		329-338 & 7kdlB & 0.75 & 1.63 & 0.96 & 0.57 & 1.98 & 1.83 & 0.71 & 1.66 & 1.37 & 1.48 & 2.52 & 2.52 \\ 
		343-348 & 6zgeC & 0.48 & 1.85 & 1.45 & 0.28 & 0.32 & 0.32 & 0.28 & 0.42 & 0.35 & 0.42 & 1.79 & 0.46 \\ 
		370-375 & 6vxxA & 0.32 & 0.44 & 0.44 & 0.13 & 0.69 & 0.31 & 0.29 & 0.36 & 0.32 & 0.26 & 2.63 & 2.36 \\ 
		370-375 & 6zgiC & 0.68 & 1.05 & 1.04 & 0.56 & 1.35 & 0.98 & 0.42 & 0.78 & 0.70 & 0.52 & 2.39 & 2.34 \\ 
		380-394 & 6x79B & 1.81 & 3.29 & 1.81 & 0.62 & 0.74 & 0.62 & 0.89 & 0.98 & 0.90 & 0.77 & 1.50 & 1.18 \\ 
		380-394 & 7kdlC & 2.38 & 3.00 & 3.00 & 0.49 & 0.61 & 0.51 & 0.80 & 1.10 & 0.80 & 0.70 & 1.04 & 0.70 \\ 
		410-416 & 6zoxB & 0.73 & 2.40 & 2.13 & 0.27 & 0.32 & 0.27 & 0.28 & 0.35 & 0.28 & 0.61 & 0.69 & 0.69 \\ 
		410-416 & 7kdkA & 0.59 & 2.44 & 2.19 & 0.41 & 0.47 & 0.43 & 0.36 & 0.69 & 0.69 & 0.37 & 1.50 & 0.48 \\ 
		422-430 & 6xm0B & 1.50 & 2.07 & 1.86 & 1.13 & 2.13 & 2.06 & 1.18 & 1.66 & 1.53 & 1.18 & 1.21 & 1.21 \\ 
		422-430 & 6xr8B & 0.96 & 1.40 & 1.06 & 0.29 & 0.34 & 0.33 & 0.46 & 0.88 & 0.65 & 0.43 & 0.43 & 0.43 \\ 
		438-451 & 6xr8A & 2.08 & 3.78 & 3.78 & 3.41 & 3.50 & 3.50 & 4.41 & 6.01 & 5.93 & 0.56 & 1.10 & 0.77 \\ 
		438-451 & 7kdlB & 2.96 & 4.81 & 4.81 & 2.47 & 6.62 & 5.15 & 2.82 & 5.81 & 4.00 & 1.89 & 8.67 & 4.19 \\ 
		454-472 & 6zgeC & 1.53 & 6.13 & 5.53 & 2.52 & 11.69 & 11.65 & 3.56 & 7.67 & 5.11 & 1.36 & 6.41 & 6.41 \\ 
		475-487 & 6xm0B & 2.31 & 13.55 & 11.37 & 1.98 & 10.64 & 2.98 & 4.16 & 13.45 & 4.69 & 3.21 & 11.59 & 8.18 \\ 
		475-487 & 7dddA & 1.41 & 12.75 & 12.75 & 0.76 & 0.97 & 0.97 & 3.37 & 4.36 & 4.36 & 2.50 & 11.35 & 4.47 \\ 
		495-506 & 6xm0B & 2.96 & 5.75 & 4.38 & 1.92 & 3.50 & 2.25 & 2.89 & 3.98 & 3.72 & 1.47 & 8.88 & 3.47 \\ 
		495-506 & 6zp0A & 2.79 & 4.92 & 4.56 & 0.95 & 7.33 & 3.40 & 1.85 & 4.39 & 4.39 & 0.57 & 0.78 & 0.71 \\ 
		495-506 & 7kdlB & 2.85 & 7.63 & 7.37 & 2.42 & 5.02 & 3.07 & 4.74 & 7.85 & 6.70 & 1.81 & 4.34 & 4.34 \\ 
		517-523 & 6xm0A & 0.76 & 1.13 & 1.02 & 0.96 & 1.41 & 1.37 & 1.03 & 1.60 & 1.40 & 0.87 & 1.19 & 1.19 \\ 
		517-523 & 6xm0B & 1.55 & 2.24 & 1.84 & 1.31 & 1.47 & 1.47 & 1.64 & 2.22 & 1.77 & 1.35 & 1.96 & 1.66 \\ 
		517-523 & 6xm3A & 0.99 & 1.44 & 1.29 & 0.80 & 1.03 & 0.92 & 0.55 & 1.87 & 1.61 & 0.79 & 1.22 & 1.00 \\ 
		517-523 & 6zoxA & 0.70 & 1.23 & 1.23 & 0.33 & 0.45 & 0.36 & 0.30 & 0.68 & 0.56 & 0.63 & 1.48 & 0.97 \\ 
		526-537 & 6x29B & 1.60 & 2.68 & 2.15 & 0.32 & 0.66 & 0.39 & 0.62 & 0.71 & 0.71 & 0.99 & 1.25 & 1.25 \\ 
		526-537 & 7ad1B & 1.68 & 1.88 & 1.88 & 0.47 & 1.14 & 0.62 & 0.50 & 0.61 & 0.50 & 1.10 & 1.92 & 1.80 \\ 
		555-564 & 7kdkC & 1.00 & 2.40 & 1.58 & 0.46 & 0.50 & 0.50 & 0.42 & 2.06 & 0.81 & 0.78 & 1.03 & 0.81 \\ 
		578-583 & 6zoxB & 0.46 & 0.98 & 0.88 & 0.20 & 0.45 & 0.41 & 0.35 & 0.75 & 0.51 & 0.28 & 0.52 & 0.41 \\ 
		600-608 & 7kdlA & 0.48 & 1.16 & 1.16 & 0.17 & 1.72 & 0.24 & 0.70 & 2.63 & 0.70 & 0.99 & 2.89 & 1.87 \\ 
		614-620 & 6xm4C & 0.58 & 2.81 & 1.76 & 0.53 & 2.08 & 2.08 & 0.52 & 0.92 & 0.92 & 0.50 & 2.84 & 0.55 \\ 
		624-641 & 6xm0B & 2.99 & 6.47 & 2.99 & 0.85 & 0.85 & 0.85 & 2.98 & 7.75 & 5.40 & 3.31 & 6.62 & 5.48 \\ 
		656-663 & 7kdkB & 0.42 & 0.88 & 0.75 & 0.33 & 0.40 & 0.36 & 0.33 & 0.41 & 0.40 & 0.46 & 1.37 & 1.10 \\ 
		697-710 & 6vxxB & 1.78 & 3.18 & 2.48 & 1.25 & 4.27 & 2.35 & 1.67 & 4.76 & 2.96 & 1.59 & 3.51 & 2.73 \\ 
		825-836 & 6xluB & 1.51 & 3.67 & 2.82 & 1.84 & 2.97 & 2.31 & 2.27 & 3.85 & 3.85 & 1.83 & 3.44 & 2.71 \\ 
		825-836 & 6xm3B & 1.48 & 5.83 & 5.61 & 0.66 & 2.08 & 0.75 & 0.86 & 1.63 & 1.62 & 1.61 & 2.22 & 2.22 \\ 
		825-836 & 6xm3C & 1.78 & 3.14 & 2.19 & 1.51 & 2.98 & 1.87 & 1.05 & 3.40 & 1.71 & 2.47 & 4.98 & 3.99 \\ 
		825-836 & 6zgiA & 1.23 & 2.66 & 2.66 & 1.39 & 2.61 & 2.61 & 1.77 & 2.54 & 2.49 & 1.54 & 3.82 & 2.07 \\ 
		841-848 & 6xluC & 0.82 & 3.97 & 2.04 & 1.21 & 3.26 & 3.13 & 1.21 & 1.91 & 1.21 & 1.06 & 1.69 & 1.31 \\ 
		841-848 & 6xm3B & 1.80 & 2.61 & 2.55 & 1.37 & 5.33 & 2.23 & 1.26 & 3.25 & 3.02 & 1.01 & 3.36 & 1.23 \\ 
		841-848 & 6xm4B & 1.39 & 5.68 & 3.12 & 1.96 & 4.09 & 2.30 & 2.45 & 4.55 & 4.55 & 0.91 & 3.25 & 3.25 \\ 
		841-848 & 6zgeB & 0.97 & 3.55 & 1.36 & 1.02 & 1.38 & 1.38 & 1.25 & 3.26 & 2.73 & 0.95 & 1.55 & 1.27 \\ 
		841-848 & 7dddB & 1.67 & 3.08 & 2.93 & 0.86 & 2.22 & 2.10 & 1.27 & 2.64 & 1.51 & 0.89 & 2.48 & 1.66 \\ 
		862-866 & 6zoxB & 0.15 & 1.60 & 0.79 & 0.13 & 0.22 & 0.15 & 0.13 & 0.23 & 0.14 & 0.20 & 0.34 & 0.34 \\ 
		891-897 & 7a4nB & 0.21 & 0.55 & 0.25 & 0.32 & 0.53 & 0.53 & 0.38 & 0.77 & 0.51 & 0.41 & 1.27 & 0.64 \\ 
		891-897 & 7kdkB & 0.39 & 0.40 & 0.40 & 0.56 & 1.26 & 1.26 & 0.33 & 0.58 & 0.58 & 0.57 & 1.25 & 1.15 \\ 
		908-913 & 7a4nB & 0.24 & 0.54 & 0.30 & 0.19 & 0.29 & 0.24 & 0.32 & 0.56 & 0.45 & 0.30 & 0.45 & 0.42 \\ 
		968-976 & 6xraC & 0.27 & 0.68 & 0.40 & 0.35 & 0.38 & 0.38 & 0.37 & 0.58 & 0.49 & 1.36 & 1.80 & 1.38 \\ 
		968-976 & 6zp0C & 0.89 & 1.15 & 1.15 & 0.20 & 0.27 & 0.20 & 0.33 & 1.70 & 0.40 & 1.20 & 2.96 & 2.19 \\ 
		1033-1046 & 7kdkA & 1.70 & 4.17 & 3.13 & 1.00 & 4.19 & 3.54 & 2.57 & 3.94 & 3.94 & 2.86 & 5.95 & 5.32 \\ 
		1106-1112 & 7kdkC & 0.71 & 1.13 & 1.09 & 0.27 & 0.36 & 0.36 & 0.57 & 1.60 & 1.22 & 0.64 & 2.01 & 0.64 \\ 
		1124-1132 & 6xm0A & 0.93 & 1.39 & 1.24 & 0.43 & 0.45 & 0.45 & 0.51 & 1.20 & 0.86 & 0.83 & 3.38 & 2.98 \\ 
		1124-1132 & 6xraC & 1.20 & 2.99 & 2.05 & 1.20 & 1.40 & 1.34 & 1.43 & 2.54 & 2.54 & 1.90 & 3.99 & 3.85 \\ 
		1135-1141 & 6xraC & 1.24 & 2.01 & 1.68 & 1.16 & 2.49 & 2.10 & 0.92 & 3.40 & 2.17 & 1.48 & 3.59 & 2.11 \\ 
		1135-1141 & 7kdkB & 0.38 & 1.08 & 0.62 & 0.13 & 0.17 & 0.17 & 0.31 & 0.38 & 0.38 & 0.38 & 1.01 & 0.61 \\ 
		\hline				
		%	\end{tabular}
		%\end{table}
	\end{longtable}

	\clearpage
	
	% latex table generated in R 3.6.3 by xtable 1.8-4 package
	% Fri Apr 30 00:30:27 2021
	\begin{longtable}[ht]{lll|rrr|rrr|rrr|rrr}
		%	\centering
		%	\begin{tabular}{lll|rrr|rrr|rrr|rrr}
		\caption{RMSD metrics for the loop instances with multiple conformations. %The four methods are assessed according to how well the decoys generated from .  
			The loop backbone RMSDs are shown for the 37 multiple conformation targets, where decoys generated from each target are compared to all known conformations for that loop instance.  
			The columns `Min.', `Top1', and `Top5' refer respectively to the lowest RMSD among the 500 decoys, RMSD of the top-ranked decoy, and lowest RMSD among the top-five ranked decoys, where each is calculated to the closest structure in the cluster represented by the chain in the `Comp.' PDB column.  The PDB column `Build' indicates the representative chain used to generate loop decoys. %, while `Comp.' indicates the representative chain containing the loop conformation to which the decoys are being compared. 		
			For example, 130--140 has three distinct loop conformations, represented by the PDB chains 6xluB, 7kdkC, and 7kdlA; using 6xluB as the input chain for generating decoys, the top five decoys of the DiSGro method included one that could predict the conformation of the 130--140 loop represented by 7kdlA with local RMSD 1.51 \AA.} \\
		\multicolumn{14}{c}{\textbf{Local RMSD}} \\
		\hline
		& \multicolumn{2}{c|}{PDB} & \multicolumn{3}{c|}{DiSGro} & \multicolumn{3}{c|}{NGK} &
		\multicolumn{3}{c|}{PETALS} & \multicolumn{3}{c}{Sphinx}   \\
		\hline
		Region & Build & Comp. & Min. & Top1 & Top5 & Min. & Top1 & Top5 & Min. & Top1 & Top5 & Min. & Top1 & Top5 \\ 
		\hline
		130-140 & 6xluB & 6xluB & 0.57 & 1.02 & 1.01 & 0.16 & 0.67 & 0.32 & 0.35 & 0.67 & 0.47 & 0.77 & 2.22 & 1.19 \\ 
		130-140 & 6xluB & 7kdkC & 1.34 & 1.56 & 1.56 & 1.08 & 1.35 & 1.17 & 0.87 & 1.42 & 1.34 & 1.17 & 2.25 & 2.07 \\ 
		130-140 & 6xluB & 7kdlA & 1.27 & 1.58 & 1.51 & 1.12 & 1.46 & 1.22 & 0.95 & 1.44 & 1.24 & 1.04 & 2.71 & 1.81 \\ 
		130-140 & 7kdkC & 6xluB & 1.35 & 2.54 & 2.37 & 0.68 & 1.96 & 1.36 & 1.39 & 2.12 & 1.86 & 1.21 & 2.09 & 1.98 \\ 
		130-140 & 7kdkC & 7kdkC & 1.64 & 3.12 & 2.67 & 0.68 & 2.47 & 1.19 & 1.32 & 2.40 & 1.91 & 1.43 & 2.38 & 2.28 \\ 
		130-140 & 7kdkC & 7kdlA & 1.52 & 2.50 & 2.44 & 1.15 & 2.18 & 1.91 & 1.57 & 2.49 & 1.93 & 1.22 & 2.81 & 2.68 \\ 
		130-140 & 7kdlA & 6xluB & 1.55 & 2.58 & 1.78 & 0.72 & 3.14 & 1.52 & 0.90 & 2.28 & 1.81 & 1.05 & 3.17 & 1.64 \\ 
		130-140 & 7kdlA & 7kdkC & 1.71 & 2.59 & 2.20 & 1.49 & 3.19 & 1.88 & 1.42 & 2.78 & 2.05 & 1.43 & 2.96 & 1.92 \\ 
		130-140 & 7kdlA & 7kdlA & 1.43 & 3.13 & 1.69 & 1.00 & 3.59 & 1.64 & 0.61 & 2.69 & 1.48 & 1.11 & 3.84 & 1.87 \\ 
		172-187 & 6zp0B & 6zp0B & 1.87 & 5.62 & 3.47 & 2.19 & 3.76 & 2.31 & 2.26 & 4.58 & 4.58 & 1.51 & 3.04 & 2.09 \\ 
		172-187 & 6zp0B & 7df3B & 2.11 & 5.28 & 3.54 & 2.19 & 4.14 & 2.45 & 1.77 & 4.81 & 4.81 & 1.71 & 2.89 & 2.40 \\ 
		172-187 & 7df3B & 6zp0B & 1.86 & 4.99 & 4.11 & 1.01 & 1.36 & 1.22 & 1.77 & 2.16 & 1.94 & 1.67 & 2.42 & 2.42 \\ 
		172-187 & 7df3B & 7df3B & 2.17 & 5.45 & 3.72 & 0.41 & 0.53 & 0.41 & 1.16 & 1.67 & 1.63 & 1.67 & 2.96 & 2.54 \\ 
		320-324 & 6zoxC & 6xm3A & 0.68 & 1.10 & 0.74 & 0.87 & 0.96 & 0.91 & 0.72 & 1.04 & 0.96 & 0.84 & 1.10 & 1.00 \\ 
		320-324 & 6zoxC & 6zoxC & 0.11 & 0.44 & 0.39 & 0.06 & 0.14 & 0.12 & 0.08 & 0.14 & 0.14 & 0.09 & 0.14 & 0.10 \\ 
		329-338 & 6x29A & 6x29A & 0.45 & 1.45 & 1.08 & 0.16 & 0.20 & 0.17 & 0.39 & 1.41 & 0.72 & 0.52 & 1.96 & 1.35 \\ 
		329-338 & 6x29A & 7kdlB & 1.58 & 2.52 & 1.83 & 1.44 & 1.56 & 1.52 & 1.34 & 2.17 & 1.53 & 1.33 & 2.67 & 2.24 \\ 
		329-338 & 7kdlB & 6x29A & 0.89 & 1.72 & 1.14 & 0.60 & 1.35 & 1.21 & 1.10 & 1.67 & 1.57 & 0.65 & 2.16 & 2.01 \\ 
		329-338 & 7kdlB & 7kdlB & 0.52 & 1.28 & 0.71 & 0.35 & 1.32 & 1.13 & 0.53 & 1.33 & 0.96 & 1.00 & 2.28 & 1.95 \\ 
		370-375 & 6vxxA & 6vxxA & 0.19 & 0.30 & 0.30 & 0.08 & 0.59 & 0.16 & 0.12 & 0.26 & 0.16 & 0.16 & 1.75 & 1.21 \\ 
		370-375 & 6vxxA & 6zgiC & 1.39 & 1.82 & 1.57 & 0.94 & 2.07 & 1.84 & 0.69 & 1.73 & 1.73 & 1.36 & 1.81 & 1.70 \\ 
		370-375 & 6zgiC & 6vxxA & 0.94 & 1.35 & 1.28 & 0.91 & 1.72 & 1.57 & 0.78 & 1.61 & 1.17 & 0.43 & 1.64 & 1.58 \\ 
		370-375 & 6zgiC & 6zgiC & 0.42 & 1.03 & 1.03 & 0.35 & 0.88 & 0.78 & 0.32 & 0.63 & 0.42 & 0.43 & 1.54 & 1.51 \\ 
		422-430 & 6xm0B & 6xm0B & 1.26 & 1.57 & 1.50 & 0.86 & 1.60 & 1.51 & 0.87 & 1.40 & 1.32 & 1.07 & 1.07 & 1.07 \\ 
		422-430 & 6xm0B & 6xr8B & 0.61 & 1.69 & 1.02 & 0.74 & 1.53 & 1.32 & 1.01 & 1.66 & 1.50 & 0.64 & 1.23 & 1.01 \\ 
		422-430 & 6xr8B & 6xm0B & 1.26 & 1.89 & 1.83 & 1.66 & 1.68 & 1.68 & 1.49 & 2.04 & 1.77 & 1.48 & 1.66 & 1.66 \\ 
		422-430 & 6xr8B & 6xr8B & 0.55 & 0.84 & 0.84 & 0.17 & 0.22 & 0.22 & 0.28 & 0.74 & 0.38 & 0.24 & 0.29 & 0.25 \\ 
		438-451 & 6xr8A & 6xr8A & 1.56 & 2.41 & 2.41 & 2.05 & 3.03 & 3.03 & 2.40 & 2.84 & 2.84 & 0.32 & 0.47 & 0.39 \\ 
		438-451 & 6xr8A & 7kdlB & 1.53 & 2.74 & 2.72 & 2.48 & 3.15 & 3.15 & 2.74 & 3.15 & 3.15 & 1.34 & 1.43 & 1.39 \\ 
		438-451 & 7kdlB & 6xr8A & 1.99 & 3.29 & 2.77 & 1.99 & 3.17 & 2.82 & 2.25 & 3.58 & 2.54 & 1.04 & 4.45 & 2.56 \\ 
		438-451 & 7kdlB & 7kdlB & 2.21 & 3.80 & 2.78 & 2.00 & 3.64 & 2.83 & 2.10 & 4.10 & 2.96 & 1.74 & 4.76 & 2.99 \\ 
		475-487 & 6xm0B & 6xm0B & 1.85 & 3.80 & 3.49 & 1.57 & 3.27 & 2.36 & 1.95 & 3.19 & 2.21 & 1.97 & 4.21 & 3.35 \\ 
		475-487 & 6xm0B & 7dddA & 1.65 & 3.18 & 2.97 & 1.37 & 3.06 & 1.89 & 1.43 & 2.61 & 2.13 & 1.47 & 3.90 & 2.78 \\ 
		475-487 & 7dddA & 6xm0B & 1.69 & 3.52 & 2.89 & 1.33 & 1.33 & 1.33 & 2.10 & 2.60 & 2.60 & 1.82 & 3.34 & 2.60 \\ 
		475-487 & 7dddA & 7dddA & 1.03 & 2.67 & 2.24 & 0.52 & 0.75 & 0.75 & 1.69 & 2.01 & 2.01 & 1.21 & 2.74 & 2.05 \\ 
		495-506 & 6xm0B & 6xm0B & 2.11 & 2.77 & 2.67 & 1.29 & 2.04 & 1.77 & 2.31 & 2.90 & 2.84 & 1.24 & 3.38 & 2.06 \\ 
		495-506 & 6xm0B & 6zp0A & 1.89 & 2.36 & 2.36 & 1.26 & 2.60 & 1.64 & 1.81 & 2.09 & 2.09 & 0.73 & 2.82 & 2.32 \\ 
		495-506 & 6xm0B & 7kdlB & 2.10 & 2.72 & 2.38 & 1.84 & 2.65 & 2.07 & 1.92 & 2.78 & 2.71 & 1.42 & 3.31 & 2.84 \\ 
		495-506 & 6zp0A & 6xm0B & 1.86 & 2.79 & 2.37 & 1.47 & 3.03 & 2.64 & 1.38 & 2.82 & 2.82 & 1.28 & 1.67 & 1.37 \\ 
		495-506 & 6zp0A & 6zp0A & 1.79 & 2.39 & 2.12 & 0.41 & 3.08 & 2.67 & 0.95 & 2.38 & 2.38 & 0.39 & 0.52 & 0.44 \\ 
		495-506 & 6zp0A & 7kdlB & 2.09 & 2.68 & 2.68 & 1.48 & 3.53 & 2.94 & 1.39 & 2.65 & 2.65 & 1.05 & 1.29 & 1.07 \\ 
		495-506 & 7kdlB & 6xm0B & 1.87 & 3.06 & 3.00 & 1.31 & 3.03 & 1.91 & 1.49 & 2.73 & 2.65 & 1.05 & 2.99 & 1.56 \\ 
		495-506 & 7kdlB & 6zp0A & 1.95 & 3.24 & 2.81 & 1.52 & 2.97 & 2.09 & 1.74 & 3.03 & 2.99 & 0.48 & 2.97 & 1.06 \\ 
		495-506 & 7kdlB & 7kdlB & 2.14 & 3.32 & 3.28 & 1.77 & 3.18 & 2.48 & 2.03 & 3.25 & 3.21 & 0.88 & 3.27 & 1.68 \\ 
		517-523 & 6xm0A & 6xm0A & 0.68 & 0.96 & 0.93 & 1.21 & 1.55 & 1.53 & 0.91 & 1.23 & 1.07 & 1.04 & 1.51 & 1.26 \\ 
		517-523 & 6xm0A & 6xm0B & 1.46 & 2.08 & 1.88 & 1.53 & 1.67 & 1.61 & 1.58 & 1.96 & 1.93 & 1.19 & 1.75 & 1.75 \\ 
		517-523 & 6xm0A & 6xm3A & 1.26 & 1.71 & 1.59 & 1.55 & 1.66 & 1.63 & 1.63 & 1.97 & 1.88 & 1.13 & 1.83 & 1.59 \\ 
		517-523 & 6xm0A & 6zoxA & 0.61 & 1.30 & 1.26 & 0.37 & 0.46 & 0.45 & 0.47 & 1.17 & 1.04 & 0.27 & 0.95 & 0.95 \\ 
		517-523 & 6xm0B & 6xm0A & 1.18 & 1.30 & 1.26 & 1.32 & 2.08 & 1.51 & 1.01 & 1.29 & 1.29 & 0.91 & 1.85 & 1.62 \\ 
		517-523 & 6xm0B & 6xm0B & 1.07 & 2.00 & 1.72 & 0.99 & 1.06 & 1.06 & 1.36 & 1.85 & 1.64 & 0.89 & 1.36 & 1.36 \\ 
		517-523 & 6xm0B & 6xm3A & 1.23 & 1.79 & 1.79 & 1.41 & 1.92 & 1.83 & 1.60 & 2.00 & 1.78 & 1.20 & 1.94 & 1.94 \\ 
		517-523 & 6xm0B & 6zoxA & 0.96 & 1.61 & 1.39 & 1.21 & 1.53 & 1.29 & 0.94 & 1.60 & 1.37 & 1.02 & 1.29 & 1.19 \\ 
		517-523 & 6xm3A & 6xm0A & 0.67 & 1.46 & 0.97 & 1.17 & 1.63 & 1.53 & 0.86 & 1.12 & 1.07 & 0.60 & 1.56 & 0.85 \\ 
		517-523 & 6xm3A & 6xm0B & 0.88 & 1.67 & 1.11 & 1.06 & 1.29 & 1.09 & 1.09 & 2.19 & 2.03 & 1.00 & 1.77 & 1.75 \\ 
		517-523 & 6xm3A & 6xm3A & 1.11 & 1.25 & 1.25 & 1.07 & 1.45 & 1.09 & 1.11 & 1.92 & 1.81 & 0.80 & 1.61 & 1.60 \\ 
		517-523 & 6xm3A & 6zoxA & 0.35 & 1.29 & 0.35 & 0.24 & 0.34 & 0.33 & 0.24 & 1.22 & 1.20 & 0.29 & 0.34 & 0.34 \\ 
		517-523 & 6zoxA & 6xm0A & 1.01 & 1.52 & 1.19 & 1.48 & 1.63 & 1.54 & 0.90 & 1.41 & 1.09 & 0.86 & 1.23 & 1.23 \\ 
		517-523 & 6zoxA & 6xm0B & 0.93 & 1.96 & 1.89 & 1.36 & 1.68 & 1.68 & 1.00 & 1.91 & 1.36 & 1.23 & 1.91 & 1.59 \\ 
		517-523 & 6zoxA & 6xm3A & 1.11 & 1.62 & 1.49 & 1.37 & 1.58 & 1.45 & 1.09 & 1.59 & 1.39 & 1.02 & 1.67 & 1.51 \\ 
		517-523 & 6zoxA & 6zoxA & 0.44 & 1.10 & 1.00 & 0.19 & 0.25 & 0.22 & 0.20 & 0.44 & 0.28 & 0.25 & 1.24 & 0.58 \\ 
		526-537 & 6x29B & 6x29B & 1.47 & 2.21 & 1.89 & 0.27 & 0.47 & 0.32 & 0.41 & 0.41 & 0.41 & 0.65 & 0.75 & 0.75 \\ 
		526-537 & 6x29B & 7ad1B & 1.58 & 2.54 & 2.24 & 0.74 & 0.95 & 0.78 & 0.80 & 0.89 & 0.89 & 0.77 & 0.97 & 0.97 \\ 
		526-537 & 7ad1B & 6x29B & 1.29 & 1.61 & 1.54 & 0.76 & 1.28 & 0.83 & 0.73 & 0.85 & 0.85 & 0.87 & 1.52 & 0.98 \\ 
		526-537 & 7ad1B & 7ad1B & 1.21 & 1.59 & 1.56 & 0.32 & 0.92 & 0.50 & 0.41 & 0.41 & 0.41 & 0.71 & 1.58 & 0.84 \\ 
		825-836 & 6xluB & 6xluB & 1.38 & 4.14 & 1.77 & 1.44 & 2.83 & 2.04 & 1.64 & 2.62 & 2.51 & 1.34 & 2.78 & 2.24 \\ 
		825-836 & 6xluB & 6xm3B & 2.03 & 3.83 & 2.39 & 2.13 & 2.30 & 2.30 & 2.34 & 2.88 & 2.88 & 1.92 & 3.04 & 2.93 \\ 
		825-836 & 6xluB & 6xm3C & 1.83 & 4.05 & 2.59 & 2.13 & 2.54 & 2.54 & 2.35 & 3.09 & 3.09 & 2.34 & 3.55 & 3.38 \\ 
		825-836 & 6xluB & 6zgiA & 1.94 & 2.64 & 2.64 & 1.83 & 3.91 & 2.13 & 2.15 & 3.40 & 3.24 & 1.90 & 2.93 & 2.85 \\ 
		825-836 & 6xm3B & 6xluB & 1.91 & 3.70 & 3.47 & 2.50 & 3.61 & 2.87 & 2.28 & 3.09 & 3.05 & 2.39 & 3.39 & 3.22 \\ 
		825-836 & 6xm3B & 6xm3B & 1.18 & 3.42 & 2.92 & 0.51 & 1.91 & 0.55 & 0.70 & 1.15 & 1.15 & 1.41 & 1.71 & 1.71 \\ 
		825-836 & 6xm3B & 6xm3C & 1.48 & 3.72 & 3.00 & 1.31 & 2.40 & 1.72 & 0.90 & 2.27 & 2.27 & 1.23 & 1.91 & 1.91 \\ 
		825-836 & 6xm3B & 6zgiA & 2.51 & 3.43 & 3.43 & 2.33 & 3.95 & 3.89 & 3.25 & 3.88 & 3.31 & 3.17 & 4.35 & 3.49 \\ 
		825-836 & 6xm3C & 6xluB & 2.06 & 2.99 & 2.99 & 2.12 & 2.99 & 2.98 & 2.47 & 3.46 & 3.13 & 2.74 & 4.16 & 3.38 \\ 
		825-836 & 6xm3C & 6xm3B & 1.41 & 2.70 & 1.47 & 0.62 & 1.91 & 1.57 & 0.95 & 1.92 & 1.42 & 1.91 & 3.10 & 2.79 \\ 
		825-836 & 6xm3C & 6xm3C & 1.48 & 2.86 & 1.98 & 1.20 & 2.35 & 1.34 & 0.75 & 2.70 & 1.44 & 1.72 & 3.72 & 3.33 \\ 
		825-836 & 6xm3C & 6zgiA & 2.24 & 3.64 & 3.47 & 2.36 & 3.76 & 3.70 & 3.22 & 3.77 & 3.54 & 2.98 & 4.97 & 3.87 \\ 
		825-836 & 6zgiA & 6xluB & 1.74 & 2.65 & 2.45 & 1.85 & 2.96 & 2.30 & 2.48 & 2.77 & 2.77 & 1.99 & 3.22 & 2.15 \\ 
		825-836 & 6zgiA & 6xm3B & 3.32 & 3.78 & 3.65 & 3.25 & 4.11 & 3.39 & 3.42 & 3.80 & 3.75 & 3.24 & 3.95 & 3.65 \\ 
		825-836 & 6zgiA & 6xm3C & 3.40 & 3.94 & 3.67 & 3.29 & 4.44 & 3.76 & 3.61 & 4.29 & 4.06 & 3.37 & 4.01 & 3.79 \\ 
		825-836 & 6zgiA & 6zgiA & 1.15 & 1.75 & 1.75 & 1.28 & 2.10 & 2.06 & 1.61 & 1.97 & 1.97 & 1.14 & 2.77 & 1.49 \\ 
		841-848 & 6xluC & 6xluC & 0.58 & 3.08 & 2.43 & 1.09 & 3.07 & 2.72 & 0.70 & 2.52 & 1.77 & 0.71 & 1.17 & 0.99 \\ 
		841-848 & 6xluC & 6xm3B & 1.11 & 1.89 & 1.80 & 1.16 & 2.08 & 2.05 & 1.31 & 2.02 & 1.92 & 0.95 & 2.82 & 2.09 \\ 
		841-848 & 6xluC & 6xm4B & 1.22 & 2.72 & 2.22 & 1.76 & 2.69 & 2.57 & 1.50 & 2.74 & 2.58 & 1.34 & 2.49 & 2.02 \\ 
		841-848 & 6xluC & 6zgeB & 0.54 & 2.60 & 0.99 & 0.37 & 2.05 & 1.94 & 0.68 & 1.43 & 0.71 & 0.42 & 0.98 & 0.45 \\ 
		841-848 & 6xluC & 7dddB & 1.69 & 2.25 & 1.91 & 1.18 & 2.32 & 1.81 & 1.30 & 2.27 & 2.03 & 1.29 & 2.71 & 2.46 \\ 
		841-848 & 6xm3B & 6xluC & 1.76 & 1.98 & 1.98 & 1.11 & 2.81 & 1.31 & 1.28 & 2.14 & 2.01 & 0.80 & 2.21 & 0.94 \\ 
		841-848 & 6xm3B & 6xm3B & 1.22 & 2.75 & 2.53 & 0.84 & 2.35 & 1.78 & 0.75 & 1.84 & 1.71 & 0.69 & 2.65 & 1.15 \\ 
		841-848 & 6xm3B & 6xm4B & 1.85 & 2.69 & 2.66 & 1.70 & 2.52 & 1.96 & 1.75 & 2.35 & 2.33 & 1.42 & 2.44 & 1.72 \\ 
		841-848 & 6xm3B & 6zgeB & 1.28 & 1.98 & 1.98 & 0.77 & 2.69 & 1.71 & 0.86 & 1.51 & 1.38 & 0.56 & 2.92 & 1.74 \\ 
		841-848 & 6xm3B & 7dddB & 0.83 & 2.09 & 2.00 & 1.66 & 2.84 & 1.97 & 1.20 & 2.72 & 2.42 & 1.39 & 2.72 & 2.38 \\ 
		841-848 & 6xm4B & 6xluC & 1.10 & 3.00 & 2.46 & 0.85 & 2.57 & 1.24 & 1.25 & 2.25 & 2.15 & 0.74 & 2.64 & 2.49 \\ 
		841-848 & 6xm4B & 6xm3B & 1.06 & 1.99 & 1.49 & 1.02 & 1.62 & 1.61 & 1.00 & 2.24 & 1.79 & 0.76 & 1.40 & 0.94 \\ 
		841-848 & 6xm4B & 6xm4B & 1.33 & 2.51 & 2.30 & 1.37 & 2.58 & 1.72 & 1.27 & 2.36 & 2.33 & 1.27 & 2.37 & 2.25 \\ 
		841-848 & 6xm4B & 6zgeB & 0.68 & 2.73 & 2.29 & 0.56 & 1.83 & 1.83 & 1.16 & 2.96 & 2.73 & 0.62 & 2.79 & 2.36 \\ 
		841-848 & 6xm4B & 7dddB & 1.21 & 2.80 & 2.33 & 1.61 & 2.75 & 2.73 & 1.47 & 2.58 & 2.51 & 1.73 & 2.36 & 2.36 \\ 
		841-848 & 6zgeB & 6xluC & 1.13 & 2.74 & 1.60 & 0.98 & 1.78 & 1.77 & 1.02 & 2.89 & 1.61 & 0.77 & 1.76 & 1.04 \\ 
		841-848 & 6zgeB & 6xm3B & 1.20 & 2.02 & 1.61 & 1.35 & 2.30 & 2.09 & 1.05 & 2.02 & 1.62 & 0.77 & 2.16 & 2.16 \\ 
		841-848 & 6zgeB & 6xm4B & 1.31 & 2.45 & 1.99 & 1.68 & 2.87 & 2.76 & 1.28 & 2.54 & 1.41 & 1.58 & 2.59 & 2.14 \\ 
		841-848 & 6zgeB & 6zgeB & 0.45 & 2.51 & 0.45 & 0.48 & 0.77 & 0.77 & 0.70 & 2.20 & 1.82 & 0.36 & 0.52 & 0.43 \\ 
		841-848 & 6zgeB & 7dddB & 1.90 & 2.64 & 2.37 & 1.34 & 2.44 & 1.42 & 1.35 & 2.20 & 2.20 & 2.10 & 2.83 & 2.65 \\ 
		841-848 & 7dddB & 6xluC & 1.80 & 2.44 & 2.44 & 1.02 & 3.13 & 2.98 & 0.95 & 2.93 & 2.06 & 0.59 & 3.12 & 2.09 \\ 
		841-848 & 7dddB & 6xm3B & 0.92 & 1.81 & 1.77 & 1.22 & 2.21 & 2.19 & 0.90 & 1.83 & 1.83 & 0.79 & 2.24 & 1.09 \\ 
		841-848 & 7dddB & 6xm4B & 2.06 & 2.66 & 2.66 & 1.80 & 3.01 & 2.99 & 1.80 & 2.66 & 2.51 & 1.38 & 2.97 & 2.04 \\ 
		841-848 & 7dddB & 6zgeB & 0.97 & 2.61 & 2.29 & 0.29 & 1.95 & 1.86 & 0.93 & 2.25 & 1.19 & 0.47 & 2.19 & 1.38 \\ 
		841-848 & 7dddB & 7dddB & 1.36 & 2.41 & 2.41 & 1.22 & 2.05 & 1.96 & 1.00 & 1.63 & 1.10 & 0.83 & 2.24 & 1.32 \\ 
		968-976 & 6xraC & 6xraC & 0.25 & 0.44 & 0.33 & 0.20 & 0.20 & 0.20 & 0.34 & 0.52 & 0.42 & 1.05 & 1.45 & 1.26 \\ 
		968-976 & 6xraC & 6zp0C & 3.09 & 3.52 & 3.43 & 1.82 & 3.52 & 3.07 & 1.50 & 3.37 & 3.37 & 0.63 & 3.69 & 3.58 \\ 
		968-976 & 6zp0C & 6xraC & 2.75 & 3.59 & 3.59 & 3.07 & 3.67 & 3.63 & 3.37 & 3.73 & 3.64 & 2.07 & 3.18 & 3.18 \\ 
		968-976 & 6zp0C & 6zp0C & 0.56 & 1.11 & 1.11 & 0.13 & 0.19 & 0.16 & 0.26 & 1.41 & 0.29 & 0.90 & 1.92 & 1.85 \\ 
		1124-1132 & 6xm0A & 6xm0A & 0.67 & 1.13 & 0.97 & 0.30 & 0.34 & 0.34 & 0.38 & 0.68 & 0.50 & 0.50 & 1.77 & 1.77 \\ 
		1124-1132 & 6xm0A & 6xraC & 2.38 & 4.44 & 2.74 & 3.47 & 3.79 & 3.79 & 2.42 & 4.12 & 3.88 & 1.91 & 2.74 & 2.70 \\ 
		1124-1132 & 6xraC & 6xm0A & 2.80 & 3.11 & 3.11 & 2.67 & 3.31 & 3.31 & 2.71 & 3.07 & 2.78 & 2.69 & 3.47 & 3.13 \\ 
		1124-1132 & 6xraC & 6xraC & 0.96 & 2.06 & 1.66 & 1.18 & 1.37 & 1.33 & 0.98 & 1.83 & 1.82 & 1.26 & 1.35 & 1.35 \\ 
		1135-1141 & 6xraC & 6xraC & 0.93 & 1.57 & 1.37 & 0.81 & 1.70 & 1.65 & 0.60 & 1.54 & 1.43 & 0.84 & 1.53 & 1.00 \\ 
		1135-1141 & 6xraC & 7kdkB & 0.55 & 1.53 & 1.19 & 0.74 & 0.94 & 0.91 & 0.61 & 1.51 & 1.32 & 0.75 & 0.85 & 0.85 \\ 
		1135-1141 & 7kdkB & 6xraC & 0.80 & 0.82 & 0.82 & 1.20 & 1.30 & 1.24 & 0.69 & 1.29 & 1.29 & 0.81 & 1.58 & 1.47 \\ 
		1135-1141 & 7kdkB & 7kdkB & 0.21 & 0.58 & 0.35 & 0.11 & 0.16 & 0.16 & 0.21 & 0.25 & 0.25 & 0.19 & 0.47 & 0.46 \\ 
		\hline 
		\multicolumn{14}{c}{~~} \\
		\multicolumn{14}{c}{\textbf{Global RMSD}} \\
		\hline
		Region & Build & Comp. & Min. & Top1 & Top5 & Min. & Top1 & Top5 & Min. & Top1 & Top5 & Min. & Top1 & Top5 \\ 
		\hline		
		130-140 & 6xluB & 6xluB & 1.08 & 1.30 & 1.23 & 0.41 & 0.87 & 0.60 & 0.63 & 0.92 & 0.64 & 1.03 & 2.54 & 1.64 \\ 
		130-140 & 6xluB & 7kdkC & 2.38 & 3.75 & 3.06 & 2.54 & 2.96 & 2.85 & 2.36 & 2.73 & 2.73 & 1.91 & 3.56 & 3.21 \\ 
		130-140 & 6xluB & 7kdlA & 1.82 & 1.94 & 1.94 & 1.57 & 1.82 & 1.67 & 1.36 & 1.86 & 1.63 & 1.69 & 3.22 & 2.69 \\ 
		130-140 & 7kdkC & 6xluB & 1.75 & 3.33 & 3.08 & 1.41 & 2.13 & 1.92 & 2.21 & 3.65 & 3.07 & 1.82 & 3.11 & 3.00 \\ 
		130-140 & 7kdkC & 7kdkC & 1.89 & 4.24 & 3.11 & 0.85 & 3.16 & 1.34 & 1.80 & 3.47 & 2.19 & 1.92 & 4.94 & 4.81 \\ 
		130-140 & 7kdkC & 7kdlA & 2.19 & 3.92 & 3.09 & 2.40 & 2.68 & 2.68 & 2.49 & 5.23 & 3.48 & 2.38 & 4.83 & 4.11 \\ 
		130-140 & 7kdlA & 6xluB & 2.10 & 3.58 & 2.48 & 1.67 & 4.76 & 2.19 & 1.36 & 5.09 & 2.82 & 1.92 & 4.24 & 1.96 \\ 
		130-140 & 7kdlA & 7kdkC & 2.71 & 3.18 & 3.18 & 2.87 & 5.70 & 4.12 & 2.49 & 6.46 & 2.49 & 2.74 & 3.98 & 3.26 \\ 
		130-140 & 7kdlA & 7kdlA & 1.63 & 5.13 & 2.10 & 1.41 & 5.40 & 3.20 & 0.83 & 5.72 & 3.88 & 1.36 & 5.66 & 2.88 \\ 
		172-187 & 6zp0B & 6zp0B & 5.09 & 12.58 & 7.80 & 3.37 & 12.41 & 3.37 & 3.66 & 11.65 & 10.39 & 5.09 & 13.69 & 8.39 \\ 
		172-187 & 6zp0B & 7df3B & 3.92 & 10.71 & 5.85 & 2.69 & 10.53 & 2.78 & 2.17 & 10.14 & 9.88 & 3.06 & 10.90 & 6.36 \\ 
		172-187 & 7df3B & 6zp0B & 5.17 & 10.86 & 9.67 & 2.66 & 3.18 & 3.05 & 3.50 & 4.66 & 4.20 & 6.09 & 10.81 & 8.53 \\ 
		172-187 & 7df3B & 7df3B & 3.12 & 9.69 & 7.78 & 0.70 & 0.80 & 0.70 & 1.36 & 2.58 & 2.20 & 3.82 & 8.48 & 7.38 \\ 
		320-324 & 6zoxC & 6xm3A & 2.12 & 2.35 & 2.35 & 2.22 & 2.64 & 2.61 & 2.21 & 2.46 & 2.34 & 2.06 & 2.40 & 2.40 \\ 
		320-324 & 6zoxC & 6zoxC & 0.15 & 0.53 & 0.53 & 0.17 & 0.29 & 0.24 & 0.14 & 0.20 & 0.20 & 0.36 & 0.57 & 0.50 \\ 
		329-338 & 6x29A & 6x29A & 0.58 & 1.50 & 1.11 & 0.36 & 0.38 & 0.38 & 0.62 & 2.05 & 0.85 & 1.04 & 3.14 & 1.89 \\ 
		329-338 & 6x29A & 7kdlB & 2.98 & 3.90 & 3.12 & 2.82 & 2.94 & 2.93 & 2.51 & 3.68 & 2.82 & 2.99 & 5.95 & 4.41 \\ 
		329-338 & 7kdlB & 6x29A & 1.26 & 2.81 & 2.17 & 1.03 & 2.39 & 1.86 & 1.70 & 2.78 & 2.14 & 1.62 & 2.52 & 2.52 \\ 
		329-338 & 7kdlB & 7kdlB & 0.75 & 1.63 & 0.96 & 0.57 & 1.98 & 1.83 & 0.71 & 1.66 & 1.37 & 1.48 & 3.82 & 2.80 \\ 
		370-375 & 6vxxA & 6vxxA & 0.32 & 0.44 & 0.44 & 0.13 & 0.69 & 0.31 & 0.29 & 0.36 & 0.32 & 0.26 & 2.63 & 2.36 \\ 
		370-375 & 6vxxA & 6zgiC & 2.25 & 3.05 & 2.67 & 2.18 & 3.36 & 3.15 & 1.98 & 3.10 & 3.04 & 2.57 & 3.99 & 3.41 \\ 
		370-375 & 6zgiC & 6vxxA & 2.04 & 2.43 & 2.31 & 2.15 & 2.47 & 2.40 & 1.90 & 2.50 & 2.30 & 1.34 & 3.90 & 3.58 \\ 
		370-375 & 6zgiC & 6zgiC & 0.68 & 1.05 & 1.04 & 0.56 & 1.35 & 0.98 & 0.42 & 0.78 & 0.70 & 0.52 & 2.39 & 2.34 \\ 
		422-430 & 6xm0B & 6xm0B & 1.50 & 2.07 & 1.86 & 1.13 & 2.13 & 2.06 & 1.18 & 1.66 & 1.53 & 1.21 & 1.21 & 1.21 \\ 
		422-430 & 6xm0B & 6xr8B & 1.69 & 2.13 & 1.96 & 1.53 & 2.54 & 2.38 & 1.68 & 2.32 & 2.01 & 1.18 & 2.03 & 1.43 \\ 
		422-430 & 6xr8B & 6xm0B & 2.19 & 2.84 & 2.67 & 2.40 & 2.40 & 2.40 & 2.18 & 2.90 & 2.46 & 2.21 & 2.52 & 2.52 \\ 
		422-430 & 6xr8B & 6xr8B & 0.96 & 1.40 & 1.06 & 0.29 & 0.34 & 0.33 & 0.46 & 0.88 & 0.65 & 0.43 & 0.43 & 0.43 \\ 
		438-451 & 6xr8A & 6xr8A & 2.08 & 3.78 & 3.78 & 3.41 & 3.50 & 3.50 & 4.41 & 6.01 & 5.93 & 0.56 & 1.10 & 0.77 \\ 
		438-451 & 6xr8A & 7kdlB & 2.96 & 4.76 & 4.76 & 4.22 & 4.41 & 4.41 & 5.17 & 6.68 & 6.36 & 2.01 & 2.59 & 2.55 \\ 
		438-451 & 7kdlB & 6xr8A & 2.96 & 4.81 & 4.81 & 2.47 & 6.62 & 5.15 & 2.82 & 5.81 & 4.00 & 1.89 & 8.67 & 4.19 \\ 
		438-451 & 7kdlB & 7kdlB & 3.66 & 5.39 & 5.31 & 3.13 & 7.37 & 5.62 & 2.88 & 6.19 & 4.41 & 2.77 & 9.96 & 4.74 \\ 
		475-487 & 6xm0B & 6xm0B & 2.58 & 13.79 & 11.47 & 1.98 & 10.83 & 3.40 & 5.76 & 13.78 & 6.25 & 4.89 & 12.10 & 9.69 \\ 
		475-487 & 6xm0B & 7dddA & 2.31 & 13.55 & 11.37 & 2.11 & 10.64 & 2.98 & 4.16 & 13.45 & 4.69 & 3.21 & 11.59 & 8.18 \\ 
		475-487 & 7dddA & 6xm0B & 2.78 & 13.03 & 13.03 & 2.16 & 2.94 & 2.94 & 5.02 & 6.82 & 6.82 & 4.38 & 12.76 & 6.59 \\ 
		475-487 & 7dddA & 7dddA & 1.41 & 12.75 & 12.75 & 0.76 & 0.97 & 0.97 & 3.37 & 4.36 & 4.36 & 2.50 & 11.35 & 4.47 \\ 
		495-506 & 6xm0B & 6xm0B & 2.96 & 5.75 & 4.38 & 2.10 & 3.50 & 2.25 & 2.95 & 4.14 & 3.74 & 1.47 & 8.88 & 3.47 \\ 
		495-506 & 6xm0B & 6zp0A & 3.11 & 5.76 & 5.22 & 1.92 & 3.56 & 2.42 & 2.89 & 3.98 & 3.72 & 2.02 & 8.95 & 4.67 \\ 
		495-506 & 6xm0B & 7kdlB & 3.86 & 5.90 & 5.57 & 2.49 & 3.81 & 3.81 & 3.79 & 4.17 & 4.17 & 3.74 & 10.37 & 6.22 \\ 
		495-506 & 6zp0A & 6xm0B & 2.79 & 4.92 & 4.85 & 2.08 & 7.52 & 3.44 & 1.85 & 4.56 & 4.56 & 1.60 & 3.36 & 3.05 \\ 
		495-506 & 6zp0A & 6zp0A & 2.87 & 5.07 & 4.56 & 0.95 & 7.33 & 3.40 & 2.05 & 4.39 & 4.39 & 0.57 & 0.78 & 0.71 \\ 
		495-506 & 6zp0A & 7kdlB & 3.63 & 7.58 & 7.58 & 3.04 & 9.56 & 5.61 & 4.69 & 7.57 & 7.57 & 3.28 & 3.48 & 3.48 \\ 
		495-506 & 7kdlB & 6xm0B & 2.85 & 7.63 & 7.46 & 2.42 & 5.02 & 3.27 & 4.74 & 7.85 & 6.70 & 1.81 & 4.34 & 4.34 \\ 
		495-506 & 7kdlB & 6zp0A & 3.64 & 7.98 & 7.37 & 2.70 & 5.98 & 3.07 & 5.02 & 8.07 & 6.70 & 1.97 & 5.15 & 5.15 \\ 
		495-506 & 7kdlB & 7kdlB & 5.25 & 10.91 & 9.37 & 2.75 & 8.02 & 3.38 & 5.99 & 10.26 & 8.95 & 3.07 & 7.56 & 7.56 \\ 
		517-523 & 6xm0A & 6xm0A & 0.76 & 1.13 & 1.02 & 1.43 & 1.65 & 1.63 & 1.03 & 1.60 & 1.40 & 1.21 & 1.59 & 1.59 \\ 
		517-523 & 6xm0A & 6xm0B & 2.54 & 4.47 & 4.05 & 2.67 & 3.18 & 3.18 & 3.04 & 4.48 & 3.75 & 1.81 & 3.84 & 3.23 \\ 
		517-523 & 6xm0A & 6xm3A & 1.45 & 3.25 & 2.85 & 1.89 & 2.03 & 2.03 & 1.92 & 3.37 & 2.62 & 1.59 & 2.78 & 2.08 \\ 
		517-523 & 6xm0A & 6zoxA & 1.25 & 1.92 & 1.60 & 0.96 & 1.41 & 1.37 & 1.03 & 1.85 & 1.75 & 0.87 & 1.19 & 1.19 \\ 
		517-523 & 6xm0B & 6xm0A & 2.20 & 2.55 & 2.49 & 2.22 & 3.92 & 2.58 & 2.22 & 2.67 & 2.67 & 1.87 & 3.85 & 3.47 \\ 
		517-523 & 6xm0B & 6xm0B & 1.55 & 2.24 & 1.84 & 1.31 & 1.47 & 1.47 & 1.64 & 2.22 & 1.77 & 1.35 & 1.96 & 1.66 \\ 
		517-523 & 6xm0B & 6xm3A & 2.18 & 2.32 & 2.32 & 2.37 & 3.19 & 2.61 & 2.21 & 2.58 & 2.56 & 2.09 & 3.35 & 2.97 \\ 
		517-523 & 6xm0B & 6zoxA & 1.68 & 2.47 & 2.21 & 1.95 & 2.35 & 2.35 & 2.15 & 2.80 & 2.42 & 1.64 & 2.80 & 2.56 \\ 
		517-523 & 6xm3A & 6xm0A & 1.29 & 1.84 & 1.64 & 1.71 & 1.72 & 1.72 & 1.30 & 1.87 & 1.61 & 0.84 & 2.00 & 1.46 \\ 
		517-523 & 6xm3A & 6xm0B & 2.30 & 2.83 & 2.53 & 2.18 & 3.08 & 2.42 & 2.30 & 4.36 & 4.28 & 1.67 & 3.95 & 3.55 \\ 
		517-523 & 6xm3A & 6xm3A & 1.40 & 1.44 & 1.44 & 1.54 & 1.93 & 1.59 & 1.40 & 3.07 & 2.90 & 1.01 & 2.74 & 2.27 \\ 
		517-523 & 6xm3A & 6zoxA & 0.99 & 1.76 & 1.29 & 0.80 & 1.03 & 0.92 & 0.55 & 1.99 & 1.76 & 0.79 & 1.22 & 1.00 \\ 
		517-523 & 6zoxA & 6xm0A & 1.36 & 1.87 & 1.51 & 1.74 & 1.95 & 1.83 & 1.14 & 1.65 & 1.48 & 1.43 & 1.89 & 1.89 \\ 
		517-523 & 6zoxA & 6xm0B & 2.43 & 3.58 & 3.39 & 3.46 & 3.97 & 3.82 & 2.98 & 4.22 & 3.69 & 2.04 & 4.05 & 3.47 \\ 
		517-523 & 6zoxA & 6xm3A & 1.73 & 2.49 & 2.18 & 2.44 & 2.84 & 2.53 & 1.84 & 2.96 & 2.51 & 1.97 & 3.02 & 2.58 \\ 
		517-523 & 6zoxA & 6zoxA & 0.70 & 1.23 & 1.23 & 0.33 & 0.45 & 0.36 & 0.30 & 0.68 & 0.56 & 0.63 & 1.48 & 0.97 \\ 
		526-537 & 6x29B & 6x29B & 1.60 & 2.68 & 2.15 & 0.32 & 0.66 & 0.39 & 0.62 & 0.71 & 0.71 & 0.99 & 1.25 & 1.25 \\ 
		526-537 & 6x29B & 7ad1B & 2.19 & 3.69 & 2.93 & 2.30 & 2.93 & 2.68 & 1.88 & 2.66 & 2.29 & 1.36 & 2.01 & 2.01 \\ 
		526-537 & 7ad1B & 6x29B & 2.54 & 2.85 & 2.66 & 2.48 & 3.05 & 2.59 & 2.30 & 2.77 & 2.73 & 2.29 & 3.84 & 3.59 \\ 
		526-537 & 7ad1B & 7ad1B & 1.68 & 1.88 & 1.88 & 0.47 & 1.14 & 0.62 & 0.50 & 0.61 & 0.50 & 1.10 & 1.92 & 1.80 \\ 
		825-836 & 6xluB & 6xluB & 1.51 & 5.41 & 2.82 & 1.84 & 2.97 & 2.31 & 2.27 & 4.24 & 4.24 & 1.83 & 3.72 & 2.71 \\ 
		825-836 & 6xluB & 6xm3B & 4.21 & 8.86 & 5.02 & 4.35 & 5.78 & 5.58 & 6.22 & 8.49 & 8.08 & 4.48 & 7.75 & 6.50 \\ 
		825-836 & 6xluB & 6xm3C & 4.56 & 9.10 & 5.06 & 4.54 & 5.63 & 5.15 & 5.87 & 8.37 & 7.97 & 5.08 & 7.61 & 6.68 \\ 
		825-836 & 6xluB & 6zgiA & 2.69 & 3.67 & 3.53 & 2.75 & 4.50 & 3.33 & 3.05 & 3.85 & 3.85 & 2.70 & 3.44 & 3.39 \\ 
		825-836 & 6xm3B & 6xluB & 4.26 & 6.72 & 6.24 & 4.08 & 6.09 & 5.38 & 5.24 & 5.70 & 5.48 & 4.43 & 6.33 & 5.53 \\ 
		825-836 & 6xm3B & 6xm3B & 1.48 & 6.01 & 5.80 & 0.66 & 2.08 & 0.75 & 0.86 & 1.63 & 1.62 & 1.61 & 2.22 & 2.22 \\ 
		825-836 & 6xm3B & 6xm3C & 1.93 & 5.83 & 5.61 & 1.88 & 3.07 & 2.48 & 2.06 & 2.84 & 2.80 & 2.72 & 4.08 & 4.08 \\ 
		825-836 & 6xm3B & 6zgiA & 4.65 & 7.66 & 6.58 & 4.75 & 8.08 & 7.50 & 6.50 & 7.66 & 7.41 & 4.62 & 8.08 & 7.18 \\ 
		825-836 & 6xm3C & 6xluB & 4.72 & 5.71 & 5.52 & 4.26 & 4.93 & 4.93 & 4.92 & 5.39 & 5.39 & 5.02 & 7.04 & 5.78 \\ 
		825-836 & 6xm3C & 6xm3B & 2.24 & 4.04 & 3.07 & 1.82 & 2.98 & 2.93 & 1.88 & 3.40 & 2.49 & 2.76 & 4.98 & 3.99 \\ 
		825-836 & 6xm3C & 6xm3C & 1.78 & 3.14 & 2.19 & 1.51 & 3.02 & 1.87 & 1.05 & 3.72 & 1.71 & 2.47 & 6.16 & 4.59 \\ 
		825-836 & 6xm3C & 6zgiA & 4.71 & 7.74 & 6.60 & 4.85 & 6.67 & 6.67 & 6.40 & 7.02 & 7.02 & 6.42 & 8.13 & 7.46 \\ 
		825-836 & 6zgiA & 6xluB & 2.99 & 3.85 & 3.85 & 2.96 & 3.78 & 2.98 & 3.38 & 3.72 & 3.72 & 3.19 & 4.59 & 3.24 \\ 
		825-836 & 6zgiA & 6xm3B & 6.09 & 7.27 & 7.27 & 5.71 & 7.30 & 6.43 & 6.77 & 6.98 & 6.98 & 5.89 & 6.82 & 6.82 \\ 
		825-836 & 6zgiA & 6xm3C & 6.21 & 7.07 & 7.07 & 5.47 & 7.25 & 6.45 & 6.90 & 7.14 & 7.14 & 5.93 & 8.08 & 7.35 \\ 
		825-836 & 6zgiA & 6zgiA & 1.23 & 2.66 & 2.66 & 1.39 & 2.61 & 2.61 & 1.77 & 2.54 & 2.49 & 1.54 & 3.82 & 2.07 \\ 
		841-848 & 6xluC & 6xluC & 0.82 & 5.94 & 3.65 & 1.57 & 4.66 & 4.45 & 1.34 & 3.35 & 2.62 & 1.07 & 1.69 & 1.69 \\ 
		841-848 & 6xluC & 6xm3B & 2.84 & 5.59 & 4.26 & 2.56 & 3.95 & 3.80 & 2.60 & 3.77 & 3.77 & 2.08 & 3.64 & 3.45 \\ 
		841-848 & 6xluC & 6xm4B & 2.85 & 6.51 & 4.66 & 3.00 & 4.61 & 4.43 & 2.97 & 4.30 & 4.28 & 2.13 & 3.53 & 3.53 \\ 
		841-848 & 6xluC & 6zgeB & 1.17 & 3.97 & 2.04 & 1.21 & 3.26 & 3.13 & 1.21 & 1.91 & 1.21 & 1.06 & 1.76 & 1.31 \\ 
		841-848 & 6xluC & 7dddB & 3.31 & 6.20 & 4.39 & 2.94 & 4.77 & 4.59 & 2.82 & 3.90 & 3.90 & 2.08 & 3.84 & 3.84 \\ 
		841-848 & 6xm3B & 6xluC & 2.78 & 3.97 & 3.87 & 2.74 & 7.40 & 3.06 & 2.69 & 4.98 & 4.55 & 2.12 & 3.47 & 3.01 \\ 
		841-848 & 6xm3B & 6xm3B & 1.80 & 3.77 & 3.54 & 1.37 & 5.33 & 2.23 & 1.26 & 3.88 & 3.02 & 1.01 & 3.73 & 1.23 \\ 
		841-848 & 6xm3B & 6xm4B & 3.49 & 4.78 & 4.63 & 2.65 & 6.49 & 2.93 & 2.63 & 4.87 & 4.21 & 2.14 & 4.61 & 2.66 \\ 
		841-848 & 6xm3B & 6zgeB & 2.35 & 2.61 & 2.55 & 2.03 & 5.84 & 3.20 & 2.18 & 3.25 & 3.23 & 2.03 & 4.19 & 3.65 \\ 
		841-848 & 6xm3B & 7dddB & 2.47 & 3.53 & 3.41 & 2.60 & 6.80 & 3.52 & 2.24 & 4.68 & 4.13 & 2.15 & 3.36 & 3.03 \\ 
		841-848 & 6xm4B & 6xluC & 2.85 & 7.43 & 4.17 & 2.40 & 5.17 & 3.15 & 2.87 & 6.20 & 6.20 & 2.19 & 3.92 & 3.92 \\ 
		841-848 & 6xm4B & 6xm3B & 2.37 & 7.32 & 4.42 & 2.59 & 4.11 & 3.08 & 2.54 & 5.18 & 5.18 & 0.91 & 3.25 & 3.25 \\ 
		841-848 & 6xm4B & 6xm4B & 1.39 & 7.82 & 4.69 & 1.96 & 4.35 & 2.30 & 2.45 & 6.50 & 6.50 & 1.79 & 4.39 & 3.81 \\ 
		841-848 & 6xm4B & 6zgeB & 2.56 & 5.68 & 3.12 & 2.53 & 4.09 & 3.91 & 3.12 & 4.55 & 4.55 & 2.70 & 3.81 & 3.81 \\ 
		841-848 & 6xm4B & 7dddB & 2.98 & 7.74 & 4.85 & 3.30 & 5.54 & 4.90 & 3.85 & 5.59 & 5.59 & 2.72 & 4.08 & 4.08 \\ 
		841-848 & 6zgeB & 6xluC & 2.27 & 5.09 & 3.04 & 1.75 & 3.07 & 3.07 & 1.55 & 4.93 & 4.30 & 1.41 & 3.49 & 2.64 \\ 
		841-848 & 6zgeB & 6xm3B & 2.93 & 4.40 & 4.28 & 2.72 & 3.98 & 3.69 & 2.24 & 4.65 & 4.18 & 2.51 & 4.37 & 3.53 \\ 
		841-848 & 6zgeB & 6xm4B & 3.89 & 5.07 & 4.59 & 3.28 & 4.54 & 4.24 & 2.89 & 5.38 & 4.78 & 2.82 & 4.68 & 3.77 \\ 
		841-848 & 6zgeB & 6zgeB & 0.97 & 3.55 & 1.36 & 1.02 & 1.38 & 1.38 & 1.25 & 3.26 & 2.73 & 0.95 & 1.55 & 1.27 \\ 
		841-848 & 6zgeB & 7dddB & 3.36 & 4.92 & 4.32 & 3.22 & 3.99 & 3.91 & 2.84 & 4.92 & 4.49 & 2.93 & 4.64 & 4.08 \\ 
		841-848 & 7dddB & 6xluC & 2.47 & 3.79 & 3.64 & 1.73 & 3.61 & 3.42 & 1.39 & 3.88 & 2.66 & 1.37 & 3.69 & 2.88 \\ 
		841-848 & 7dddB & 6xm3B & 1.75 & 3.76 & 3.67 & 2.47 & 3.75 & 3.63 & 1.56 & 3.35 & 2.86 & 1.15 & 3.59 & 2.43 \\ 
		841-848 & 7dddB & 6xm4B & 4.93 & 6.26 & 6.18 & 3.70 & 4.94 & 4.83 & 3.93 & 4.94 & 4.45 & 3.15 & 4.83 & 4.46 \\ 
		841-848 & 7dddB & 6zgeB & 1.67 & 3.08 & 2.93 & 0.86 & 2.22 & 2.10 & 1.27 & 2.64 & 1.51 & 0.89 & 2.48 & 1.81 \\ 
		841-848 & 7dddB & 7dddB & 2.77 & 4.22 & 4.01 & 1.51 & 3.43 & 3.22 & 1.50 & 3.07 & 2.26 & 1.34 & 3.18 & 1.66 \\ 
		968-976 & 6xraC & 6xraC & 0.27 & 0.68 & 0.40 & 0.35 & 0.38 & 0.38 & 0.37 & 0.58 & 0.49 & 1.36 & 1.80 & 1.38 \\ 
		968-976 & 6xraC & 6zp0C & 6.50 & 7.00 & 6.81 & 3.67 & 6.88 & 5.03 & 4.60 & 7.01 & 6.27 & 2.76 & 6.22 & 6.13 \\ 
		968-976 & 6zp0C & 6xraC & 5.96 & 7.14 & 6.97 & 5.27 & 7.25 & 7.21 & 6.84 & 7.71 & 7.13 & 5.91 & 7.04 & 6.62 \\ 
		968-976 & 6zp0C & 6zp0C & 0.89 & 1.15 & 1.15 & 0.20 & 0.27 & 0.20 & 0.33 & 1.70 & 0.40 & 1.20 & 2.96 & 2.19 \\ 
		1124-1132 & 6xm0A & 6xm0A & 0.93 & 1.39 & 1.24 & 0.43 & 0.45 & 0.45 & 0.51 & 1.20 & 0.86 & 0.83 & 3.38 & 2.98 \\ 
		1124-1132 & 6xm0A & 6xraC & 4.84 & 6.84 & 6.09 & 6.45 & 7.05 & 7.05 & 5.22 & 6.99 & 6.78 & 4.79 & 6.58 & 6.12 \\ 
		1124-1132 & 6xraC & 6xm0A & 5.93 & 7.71 & 7.49 & 5.79 & 6.79 & 6.76 & 6.29 & 7.26 & 7.24 & 5.90 & 7.17 & 7.17 \\ 
		1124-1132 & 6xraC & 6xraC & 1.20 & 2.99 & 2.05 & 1.20 & 1.40 & 1.34 & 1.43 & 2.54 & 2.54 & 1.90 & 3.99 & 3.85 \\ 
		1135-1141 & 6xraC & 6xraC & 1.24 & 2.01 & 1.68 & 1.16 & 2.49 & 2.10 & 0.92 & 3.40 & 2.17 & 1.48 & 3.59 & 2.11 \\ 
		1135-1141 & 6xraC & 7kdkB & 3.32 & 4.33 & 4.33 & 3.35 & 3.89 & 3.89 & 3.26 & 4.39 & 3.73 & 3.00 & 3.74 & 3.65 \\ 
		1135-1141 & 7kdkB & 6xraC & 4.63 & 5.66 & 5.31 & 5.57 & 5.70 & 5.58 & 4.95 & 5.70 & 5.67 & 3.15 & 5.53 & 5.40 \\ 
		1135-1141 & 7kdkB & 7kdkB & 0.38 & 1.08 & 0.62 & 0.13 & 0.17 & 0.17 & 0.31 & 0.38 & 0.38 & 0.38 & 1.01 & 0.61 \\ 		
		\hline
		%	\end{tabular}
	\end{longtable}
	\begin{table}[ht]
		\footnotesize
		\caption{RMSD metrics for the loop regions with sequence variants.
			The loop backbone RMSDs are shown for each of these four loop regions, where decoys generated from each target are compared to all known sequence variants for that loop region.
			The columns `Min.', `Top1', and `Top5' refer respectively to the lowest RMSD among the 500 decoys, RMSD of the top-ranked decoy, and lowest RMSD among the top-five ranked decoys, where each is calculated to the closest structure in the cluster represented by the chain in the `Comp.' PDB column.  The PDB column `Build' indicates the representative chain used to generate loop decoys. %, while `Comp.' indicates the representative chain containing the loop conformation to which the decoys are being compared. 		 
			For example, 380--394 two different residue sequences in the PDB, represented by the PDB chains 6x79B and 7kdlC; using 6x79B as the input chain for generating decoys, the top decoy of the NGK method could predict the conformation of the 380--394 loop in 7kdlC (which had the S383C mutation) with global RMSD 0.59 \AA. \strut}
		\centering
		
		\textbf{Local RMSD}	
		\begin{tabular}{lll|rrr|rrr|rrr|rrr}
			\hline
			& \multicolumn{2}{c|}{PDB} & \multicolumn{3}{c|}{DiSGro} & \multicolumn{3}{c|}{NGK} &
			\multicolumn{3}{c|}{PETALS} & \multicolumn{3}{c}{Sphinx}   \\
			\hline
			Region & Build & Comp. & Min. & Top1 & Top5 & Min. & Top1 & Top5 & Min. & Top1 & Top5 & Min. & Top1 & Top5 \\ 
			\hline
			380-394 & 6x79B & 6x79B & 1.45 & 3.17 & 1.45 & 0.53 & 0.58 & 0.55 & 0.64 & 0.72 & 0.72 & 0.63 & 0.70 & 0.70 \\ 
			380-394 & 6x79B & 7kdlC & 1.22 & 2.96 & 1.22 & 0.40 & 0.44 & 0.44 & 0.64 & 0.71 & 0.70 & 0.44 & 0.59 & 0.46 \\ 
			380-394 & 7kdlC & 6x79B & 2.14 & 2.97 & 2.97 & 0.41 & 0.49 & 0.47 & 0.71 & 1.07 & 0.71 & 0.69 & 0.88 & 0.74 \\ 
			380-394 & 7kdlC & 7kdlC & 2.06 & 2.74 & 2.74 & 0.41 & 0.49 & 0.43 & 0.57 & 0.90 & 0.57 & 0.42 & 0.55 & 0.52 \\ 
			410-416 & 6zoxB & 6zoxB & 0.71 & 1.88 & 1.76 & 0.18 & 0.20 & 0.18 & 0.20 & 0.32 & 0.20 & 0.35 & 0.62 & 0.62 \\ 
			410-416 & 6zoxB & 7kdkA & 0.33 & 1.61 & 1.53 & 0.13 & 0.13 & 0.13 & 0.25 & 0.30 & 0.26 & 0.32 & 0.45 & 0.35 \\ 
			410-416 & 7kdkA & 6zoxB & 0.68 & 1.82 & 1.79 & 0.32 & 0.37 & 0.37 & 0.22 & 0.26 & 0.26 & 0.21 & 1.08 & 0.34 \\ 
			410-416 & 7kdkA & 7kdkA & 0.28 & 1.49 & 1.46 & 0.17 & 0.22 & 0.17 & 0.15 & 0.25 & 0.25 & 0.08 & 0.87 & 0.19 \\ 
			891-897 & 7a4nB & 7a4nB & 0.16 & 0.47 & 0.20 & 0.29 & 0.53 & 0.52 & 0.26 & 0.55 & 0.43 & 0.23 & 1.02 & 0.46 \\ 
			891-897 & 7a4nB & 7kdkB & 0.17 & 0.47 & 0.20 & 0.18 & 0.44 & 0.42 & 0.24 & 0.49 & 0.33 & 0.21 & 0.95 & 0.39 \\ 
			891-897 & 7kdkB & 7a4nB & 0.34 & 0.34 & 0.34 & 0.51 & 0.98 & 0.96 & 0.27 & 0.36 & 0.36 & 0.45 & 1.18 & 1.00 \\ 
			891-897 & 7kdkB & 7kdkB & 0.27 & 0.29 & 0.27 & 0.42 & 0.83 & 0.82 & 0.22 & 0.33 & 0.33 & 0.32 & 0.97 & 0.76 \\ 
			\hline
		\end{tabular}
		\vspace{0.1in}
		
		\textbf{Global RMSD}
		\begin{tabular}{lll|rrr|rrr|rrr|rrr}
			\hline
			& \multicolumn{2}{c|}{Cluster} & \multicolumn{3}{c|}{DiSGro} & \multicolumn{3}{c|}{NGK} &
			\multicolumn{3}{c|}{PETALS} & \multicolumn{3}{c}{Sphinx}   \\
			\hline
			Region & Build & Comp. & Min. & Top1 & Top5 & Min. & Top1 & Top5 & Min. & Top1 & Top5 & Min. & Top1 & Top5 \\ 		
			\hline
			380-394 & 6x79B & 6x79B & 1.81 & 3.29 & 1.81 & 0.62 & 0.74 & 0.62 & 0.89 & 0.98 & 0.90 & 0.77 & 1.50 & 1.18 \\ 
			380-394 & 6x79B & 7kdlC & 1.43 & 3.06 & 1.43 & 0.47 & 0.59 & 0.47 & 0.76 & 0.83 & 0.76 & 0.64 & 1.07 & 0.79 \\ 
			380-394 & 7kdlC & 6x79B & 2.38 & 3.13 & 3.13 & 0.48 & 0.62 & 0.57 & 0.88 & 1.28 & 0.88 & 0.88 & 1.38 & 0.88 \\ 
			380-394 & 7kdlC & 7kdlC & 2.38 & 3.00 & 3.00 & 0.49 & 0.61 & 0.51 & 0.80 & 1.10 & 0.80 & 0.70 & 1.04 & 0.70 \\ 
			410-416 & 6zoxB & 6zoxB & 0.73 & 2.40 & 2.13 & 0.27 & 0.32 & 0.27 & 0.28 & 0.35 & 0.28 & 0.61 & 0.69 & 0.69 \\ 
			410-416 & 6zoxB & 7kdkA & 0.47 & 2.28 & 2.08 & 0.27 & 0.27 & 0.27 & 0.32 & 0.43 & 0.32 & 0.77 & 0.79 & 0.79 \\ 
			410-416 & 7kdkA & 6zoxB & 0.82 & 2.85 & 2.39 & 0.56 & 0.63 & 0.61 & 0.46 & 0.72 & 0.72 & 0.32 & 1.70 & 0.67 \\ 
			410-416 & 7kdkA & 7kdkA & 0.59 & 2.44 & 2.19 & 0.41 & 0.47 & 0.43 & 0.36 & 0.69 & 0.69 & 0.37 & 1.50 & 0.48 \\ 
			891-897 & 7a4nB & 7a4nB & 0.21 & 0.55 & 0.25 & 0.32 & 0.53 & 0.53 & 0.38 & 0.77 & 0.51 & 0.41 & 1.27 & 0.64 \\ 
			891-897 & 7a4nB & 7kdkB & 0.27 & 0.52 & 0.27 & 0.33 & 0.57 & 0.55 & 0.39 & 0.78 & 0.53 & 0.40 & 1.17 & 0.55 \\ 
			891-897 & 7kdkB & 7a4nB & 0.39 & 0.39 & 0.39 & 0.58 & 1.48 & 1.47 & 0.35 & 0.56 & 0.56 & 0.63 & 1.58 & 1.45 \\ 
			891-897 & 7kdkB & 7kdkB & 0.39 & 0.40 & 0.40 & 0.56 & 1.26 & 1.26 & 0.33 & 0.58 & 0.58 & 0.57 & 1.25 & 1.15 \\ 
			\hline
		\end{tabular}	
	\end{table}

	% latex table generated in R 3.6.3 by xtable 1.8-4 package
	% Fri Apr 30 17:33:44 2021
	\begin{table}[ht]
		\footnotesize
		\caption{RMSD metrics for the loop targets omitted from the main analysis, as one or more methods were unsuccessful at decoy generation.
			The loop backbone RMSDs are shown for these 5 targets, where decoys generated from each target are compared to all known conformations for that loop instance.
			The columns `Min.', `Top1', and `Top5' refer respectively to the lowest RMSD among the 500 decoys, RMSD of the top-ranked decoy, and lowest RMSD among the top-five ranked decoys, where each is calculated to the closest structure in the cluster represented by the chain in the `Comp.' PDB column.  The PDB column `Build' indicates the representative chain used to generate loop decoys. %, while `Comp.' indicates the representative chain containing the loop conformation to which the decoys are being compared. 
			The dash `---' indicates that a method could not generate decoys for that target. \strut}	
		\centering
		\textbf{Local RMSD}
		\begin{tabular}{lll|rrr|rrr|rrr|rrr}
			\hline
			& \multicolumn{2}{c|}{Cluster} & \multicolumn{3}{c|}{DiSGro} & \multicolumn{3}{c|}{NGK} &
			\multicolumn{3}{c|}{PETALS} & \multicolumn{3}{c}{Sphinx}   \\
			\hline
			Region & Build & Comp. & Min. & Top1 & Top5 & Min. & Top1 & Top5 & Min. & Top1 & Top5 & Min. & Top1 & Top5 \\ 		
			\hline
			31-46 & 7a4nB & 7a4nB & 1.46 & 2.07 & 1.87 & 1.45 & 1.62 & 1.58 & 1.56 & 1.95 & 1.88 & --- & --- & --- \\ 
			146-168 & 6zgiB & 6zgiB & --- & --- & --- & 1.77 & 2.47 & 1.77 & 1.39 & 1.71 & 1.58 & --- & --- & --- \\ 
			146-168 & 6zgiB & 7dddC & --- & --- & --- & 1.78 & 2.98 & 1.78 & 1.54 & 1.91 & 1.79 & --- & --- & --- \\ 
			146-168 & 7dddC & 6zgiB & --- & --- & --- & 2.12 & 2.49 & 2.49 & 2.35 & 3.30 & 2.94 & --- & --- & --- \\ 
			146-168 & 7dddC & 7dddC & --- & --- & --- & 2.09 & 2.13 & 2.13 & 1.80 & 2.73 & 2.51 & --- & --- & --- \\ 
			320-324 & 6xm3A & 6xm3A & 0.11 & 0.44 & 0.44 & 0.12 & 0.50 & 0.37 & 0.09 & 0.34 & 0.19 & --- & --- & --- \\ 
			320-324 & 6xm3A & 6zoxC & 0.28 & 0.42 & 0.42 & 0.43 & 0.52 & 0.44 & 0.30 & 0.54 & 0.43 & --- & --- & --- \\ 
			783-816 & 6zp0C & 6zp0C & --- & --- & --- & 2.83 & 7.03 & 3.31 & 5.35 & 7.41 & 7.41 & --- & --- & --- \\ 
			\hline
		\end{tabular}
		\vspace{0.1in}
		
		\textbf{Global RMSD}
		\begin{tabular}{lll|rrr|rrr|rrr|rrr}
			\hline
			& \multicolumn{2}{c|}{Cluster} & \multicolumn{3}{c|}{DiSGro} & \multicolumn{3}{c|}{NGK} &
			\multicolumn{3}{c|}{PETALS} & \multicolumn{3}{c}{Sphinx}   \\
			\hline
			Region & Build & Comp. & Min. & Top1 & Top5 & Min. & Top1 & Top5 & Min. & Top1 & Top5 & Min. & Top1 & Top5 \\ 		
			\hline
			31-46 & 7a4nB & 7a4nB & 1.85 & 2.54 & 2.08 & 1.66 & 1.92 & 1.86 & 2.13 & 2.80 & 2.60 & --- & --- & --- \\ 
			146-168 & 6zgiB & 6zgiB & --- & --- & --- & 2.28 & 2.80 & 2.28 & 1.61 & 2.18 & 1.62 & --- & --- & --- \\ 
			146-168 & 6zgiB & 7dddC & --- & --- & --- & 2.50 & 3.35 & 2.73 & 2.17 & 2.39 & 2.20 & --- & --- & --- \\ 
			146-168 & 7dddC & 6zgiB & --- & --- & --- & 2.79 & 3.03 & 3.03 & 2.85 & 4.27 & 3.64 & --- & --- & --- \\ 
			146-168 & 7dddC & 7dddC & --- & --- & --- & 2.45 & 2.45 & 2.45 & 2.24 & 3.71 & 3.15 & --- & --- & --- \\ 
			320-324 & 6xm3A & 6xm3A & 0.20 & 0.65 & 0.50 & 0.21 & 0.97 & 0.71 & 0.24 & 0.50 & 0.47 & --- & --- & --- \\ 
			320-324 & 6xm3A & 6zoxC & 1.96 & 2.21 & 2.13 & 2.06 & 2.06 & 2.06 & 2.09 & 2.20 & 2.20 & --- & --- & --- \\ 
			783-816 & 6zp0C & 6zp0C & --- & --- & --- & 3.48 & 26.78 & 3.91 & 10.38 & 11.99 & 11.92 & --- & --- & --- \\ 
			\hline
		\end{tabular}	
	\end{table}


\begin{thebibliography}{}
	
	\bibitem[Ali and Vijayan, 2020]{ali2020dynamics}
	Ali, A. and Vijayan, R. (2020).
	\newblock Dynamics of the ACE2--SARS-CoV-2/SARS-CoV spike protein interface
	reveal unique mechanisms.
	\newblock {\em Scientific reports}, 10, 14214.
	
	\bibitem[Barozet et~al., 2021]{barozet2021protein}
	Barozet, A., Bianciotto, M., Vaisset, M., Simeon, T., Minoux, H., and
	Cort{\'e}s, J. (2021).
	\newblock Protein loops with multiple meta-stable conformations: A challenge
	for sampling and scoring methods.
	\newblock {\em Proteins: Structure, Function, and Bioinformatics},
	89(2):218--231.
	
	\bibitem[Berman et~al., 2000]{berman2000protein}
	Berman, H.~M., Westbrook, J., Feng, Z., Gilliland, G., Bhat, T.~N., Weissig,
	H., Shindyalov, I.~N., and Bourne, P.~E. (2000).
	\newblock The protein data bank.
	\newblock {\em Nucleic acids research}, 28(1):235--242.
	
	\bibitem[Cai et~al., 2020]{cai2020distinct}
	Cai, Y., Zhang, J., Xiao, T., Peng, H., Sterling, S.~M., Walsh, R.~M., Rawson,
	S., Rits-Volloch, S., and Chen, B. (2020).
	\newblock Distinct conformational states of SARS-CoV-2 spike protein.
	\newblock {\em Science}, 369(6511):1586--1592.
	
	\bibitem[Chen et~al., 2020]{chen2020mutations}
	Chen, J., Wang, R., Wang, M., and Wei, G.-W. (2020).
	\newblock Mutations strengthened SARS-CoV-2 infectivity.
	\newblock {\em Journal of Molecular Biology}, 432(19):5212--5226.
	
	\bibitem[Choi and Deane, 2010]{choi2010fread}
	Choi, Y. and Deane, C.~M. (2010).
	\newblock FREAD revisited: accurate loop structure prediction using a database
	search algorithm.
	\newblock {\em Proteins: Structure, Function, and Bioinformatics},
	78(6):1431--1440.
	
	\bibitem[Dong et~al., 2013]{dong2013optimized}
	Dong, G.~Q., Fan, H., Schneidman-Duhovny, D., Webb, B., and Sali, A. (2013).
	\newblock Optimized atomic statistical potentials: assessment of protein
	interfaces and loops.
	\newblock {\em Bioinformatics}, 29(24):3158--3166.
	
	\bibitem[Dunbar et~al., 2016]{dunbar2016sabpred}
	Dunbar, J., Krawczyk, K., Leem, J., Marks, C., Nowak, J., Regep, C., Georges,
	G., Kelm, S., Popovic, B., and Deane, C.~M. (2016).
	\newblock Sabpred: a structure-based antibody prediction server.
	\newblock {\em Nucleic acids research}, 44(W1):W474--W478.
	
	\bibitem[Espadaler et~al., 2006]{espadaler2006identification}
	Espadaler, J., Querol, E., Aviles, F.~X., and Oliva, B. (2006).
	\newblock Identification of function-associated loop motifs and application to
	protein function prediction.
	\newblock {\em Bioinformatics}, 22(18):2237--2243.
	
	\bibitem[Fiser et~al., 2000]{fiser2000modeling}
	Fiser, A., Do, R. K.~G., and {\v{S}}ali, A. (2000).
	\newblock Modeling of loops in protein structures.
	\newblock {\em Protein science}, 9(9):1753--1773.
	
	\bibitem[Grubaugh et~al., 2020]{grubaugh2020making}
	Grubaugh, N.~D., Hanage, W.~P., and Rasmussen, A.~L. (2020).
	\newblock Making sense of mutation: what D614G means for the COVID-19 pandemic
	remains unclear.
	\newblock {\em Cell}, 182(4):794--795.
	
	\bibitem[Guo et~al., 2021]{guo2021engineered}
	Guo, L., Bi, W., Wang, X., Xu, W., Yan, R., Zhang, Y., Zhao, K., Li, Y., Zhang,
	M., Cai, X., et~al. (2021).
	\newblock Engineered trimeric ACE2 binds viral spike protein and locks it in
	“three-up” conformation to potently inhibit SARS-CoV-2 infection.
	\newblock {\em Cell research}, 31(1):98--100.
	
	\bibitem[Heffernan et~al., 2017]{heffernan2017capturing}
	Heffernan, R., Yang, Y., Paliwal, K., and Zhou, Y. (2017).
	\newblock Capturing non-local interactions by long short-term memory
	bidirectional recurrent neural networks for improving prediction of protein
	secondary structure, backbone angles, contact numbers and solvent
	accessibility.
	\newblock {\em Bioinformatics}, 33(18):2842--2849.
	
	\bibitem[Henzler-Wildman and Kern, 2007]{henzler2007dynamic}
	Henzler-Wildman, K. and Kern, D. (2007).
	\newblock Dynamic personalities of proteins.
	\newblock {\em Nature}, 450(7172):964--972.
	
	\bibitem[Jiang et~al., 2020]{jiang2020neutralizing}
	Jiang, S., Hillyer, C., and Du, L. (2020).
	\newblock Neutralizing antibodies against SARS-CoV-2 and other human
	coronaviruses.
	\newblock {\em Trends in immunology}, 41(5):355--359.
	
	\bibitem[Kabsch, 1976]{kabsch1976solution}
	Kabsch, W. (1976).
	\newblock A solution for the best rotation to relate two sets of vectors.
	\newblock {\em Acta Crystallographica Section A: Crystal Physics, Diffraction,
		Theoretical and General Crystallography}, 32(5):922--923.
	
	\bibitem[Kabsch and Sander, 1983]{Kabsch1983}
	Kabsch, W. and Sander, C. (1983).
	\newblock Dictionary of protein secondary structure - pattern-recognition of
	hydrogen-bonded and geometrical features.
	\newblock {\em Biopolymers}, 22(12):2577--2637.
	
	\bibitem[Karami et~al., 2019]{karami2019dareus}
	Karami, Y., Rey, J., Postic, G., Murail, S., Tuff{\'e}ry, P., and De~Vries,
	S.~J. (2019).
	\newblock Dareus-loop: a web server to model multiple loops in homology models.
	\newblock {\em Nucleic acids research}, 47(W1):W423--W428.
	
	\bibitem[Lan et~al., 2020]{lan2020structure}
	Lan, J., Ge, J., Yu, J., Shan, S., Zhou, H., Fan, S., Zhang, Q., Shi, X., Wang,
	Q., Zhang, L., et~al. (2020).
	\newblock Structure of the SARS-CoV-2 spike receptor-binding domain bound to
	the ACE2 receptor.
	\newblock {\em Nature}, 581(7807):215--220.
	
	\bibitem[Li et~al., 2011]{li2011vsgb}
	Li, J., Abel, R., Zhu, K., Cao, Y., Zhao, S., and Friesner, R.~A. (2011).
	\newblock The VSGB 2.0 model: a next generation energy model for high
	resolution protein structure modeling.
	\newblock {\em Proteins: Structure, Function, and Bioinformatics},
	79(10):2794--2812.
	
	\bibitem[Li et~al., 2020]{LiQ}
	Li, Q., Wu, J., Nie, J., Zhang, L., Hao, H., Liu, S., et~al. (2020).
	\newblock The impact of mutations in SARS-CoV-2 spike on viral infectivity and
	antigenicity.
	\newblock {\em Cell}, 182(5):1284--1294.
	
	\bibitem[Liang et~al., 2014]{liang2014leap}
	Liang, S., Zhang, C., and Zhou, Y. (2014).
	\newblock Leap: Highly accurate prediction of protein loop conformations by
	integrating coarse-grained sampling and optimized energy scores with all-atom
	refinement of backbone and side chains.
	\newblock {\em Journal of computational chemistry}, 35(4):335--341.
	
	\bibitem[Linding et~al., 2003]{linding2003protein}
	Linding, R., Jensen, L.~J., Diella, F., Bork, P., Gibson, T.~J., and Russell,
	R.~B. (2003).
	\newblock Protein disorder prediction: implications for structural proteomics.
	\newblock {\em Structure}, 11(11):1453--1459.
	
	\bibitem[Marks et~al., 2017]{marks2017sphinx}
	Marks, C., Nowak, J., Klostermann, S., Georges, G., Dunbar, J., Shi, J., Kelm,
	S., and Deane, C.~M. (2017).
	\newblock Sphinx: merging knowledge-based and ab initio approaches to improve
	protein loop prediction.
	\newblock {\em Bioinformatics}, 33(9):1346--1353.
	
	\bibitem[Marks et~al., 2018]{marks2018predicting}
	Marks, C., Shi, J., and Deane, C.~M. (2018).
	\newblock Predicting loop conformational ensembles.
	\newblock {\em Bioinformatics}, 34(6):949--956.
	
	\bibitem[Miranda, 2018]{pyswarmsJOSS2018}
	Miranda, L. J.~V. (2018).
	\newblock {P}y{S}warms, a research-toolkit for {P}article {S}warm
	{O}ptimization in {P}ython.
	\newblock {\em Journal of Open Source Software}, 3(21), 433.
	
	\bibitem[Mittermaier and Kay, 2006]{mittermaier2006new}
	Mittermaier, A. and Kay, L.~E. (2006).
	\newblock New tools provide new insights in nmr studies of protein dynamics.
	\newblock {\em Science}, 312(5771):224--228.
	
	\bibitem[Muhammed and Aki-Yalcin, 2019]{muhammed2019homology}
	Muhammed, M.~T. and Aki-Yalcin, E. (2019).
	\newblock Homology modeling in drug discovery: Overview, current applications,
	and future perspectives.
	\newblock {\em Chemical biology \& drug design}, 93(1):12--20.
	
	\bibitem[Papaleo et~al., 2016]{papaleo2016role}
	Papaleo, E., Saladino, G., Lambrughi, M., Lindorff-Larsen, K., Gervasio, F.~L.,
	and Nussinov, R. (2016).
	\newblock The role of protein loops and linkers in conformational dynamics and
	allostery.
	\newblock {\em Chemical reviews}, 116(11):6391--6423.
	
	\bibitem[Polack et~al., 2020]{polack2020safety}
	Polack, F.~P., Thomas, S.~J., Kitchin, N., Absalon, J., Gurtman, A., Lockhart,
	S., Perez, J.~L., P{\'e}rez~Marc, G., Moreira, E.~D., Zerbini, C., et~al.
	(2020).
	\newblock Safety and efficacy of the bnt162b2 mRNA COVID-19 vaccine.
	\newblock {\em New England Journal of Medicine}, 383(27):2603--2615.
	
	\bibitem[Schneider et~al., 2014]{schneider2014local}
	Schneider, B., Gelly, J.-C., de~Brevern, A.~G., and {\v{C}}ern{\`y}, J. (2014).
	\newblock Local dynamics of proteins and dna evaluated from crystallographic B
	factors.
	\newblock {\em Acta Crystallographica Section D: Biological Crystallography},
	70(9):2413--2419.
	
	\bibitem[Schoof et~al., 2020]{schoof2020ultrapotent}
	Schoof, M., Faust, B., Saunders, R.~A., Sangwan, S., Rezelj, V., Hoppe, N.,
	Boone, M., Billesb{\o}lle, C.~B., Puchades, C., Azumaya, C.~M., et~al.
	(2020).
	\newblock An ultrapotent synthetic nanobody neutralizes SARS-CoV-2 by
	stabilizing inactive spike.
	\newblock {\em Science}, 370(6523):1473--1479.
	
	\bibitem[Sedova et~al., 2020]{sedova2020coronavirus3d}
	Sedova, M., Jaroszewski, L., Alisoltani, A., and Godzik, A. (2020).
	\newblock Coronavirus3d: 3d structural visualization of COVID-19 genomic
	divergence.
	\newblock {\em Bioinformatics}, 36(15):4360--4362.
	
	\bibitem[Sewell et~al., 2020]{sewell2020covid}
	Sewell, H.~F., Agius, R.~M., Kendrick, D., and Stewart, M. (2020).
	\newblock COVID-19 vaccines: delivering protective immunity.
	\newblock {\em BMJ}, 371:m4838.
	
	\bibitem[Shang et~al., 2020]{shang2020structural}
	Shang, J., Ye, G., Shi, K., Wan, Y., Luo, C., Aihara, H., Geng, Q., Auerbach,
	A., and Li, F. (2020).
	\newblock Structural basis of receptor recognition by SARS-CoV-2.
	\newblock {\em Nature}, 581(7807):221--224.
	
	\bibitem[Shehu et~al., 2006]{shehu2006modeling}
	Shehu, A., Clementi, C., and Kavraki, L.~E. (2006).
	\newblock Modeling protein conformational ensembles: from missing loops to
	equilibrium fluctuations.
	\newblock {\em Proteins: Structure, Function, and Bioinformatics},
	65(1):164--179.
	
	\bibitem[Shi et~al., 2020]{shi2020human}
	Shi, R., Shan, C., Duan, X., Chen, Z., Liu, P., Song, J., Song, T., Bi, X.,
	Han, C., Wu, L., et~al. (2020).
	\newblock A human neutralizing antibody targets the receptor-binding site of
	SARS-CoV-2.
	\newblock {\em Nature}, 584(7819):120--124.
	
	\bibitem[Sokal, 1958]{sokal1958statistical}
	Sokal, R.~R. (1958).
	\newblock A statistical method for evaluating systematic relationships.
	\newblock {\em Univ. Kansas, Sci. Bull.}, 38:1409--1438.
	
	\bibitem[Soto et~al., 2008]{soto2008loop}
	Soto, C.~S., Fasnacht, M., Zhu, J., Forrest, L., and Honig, B. (2008).
	\newblock Loop modeling: Sampling, filtering, and scoring.
	\newblock {\em Proteins: Structure, Function, and Bioinformatics},
	70(3):834--843.
	
	\bibitem[Stein and Kortemme, 2013]{stein2013improvements}
	Stein, A. and Kortemme, T. (2013).
	\newblock Improvements to robotics-inspired conformational sampling in rosetta.
	\newblock {\em PloS one}, 8(5): e63090.
	
	\bibitem[Tang et~al., 2014]{tang2014fast}
	Tang, K., Zhang, J., and Liang, J. (2014).
	\newblock Fast protein loop sampling and structure prediction using
	distance-guided sequential chain-growth monte carlo method.
	\newblock {\em PLoS computational biology}, 10:e1003539.
	
	\bibitem[Wang and Dunbrack~Jr, 2003]{wang2003pisces}
	Wang, G. and Dunbrack~Jr, R.~L. (2003).
	\newblock Pisces: a protein sequence culling server.
	\newblock {\em Bioinformatics}, 19(12):1589--1591.
	
	\bibitem[Waterhouse et~al., 2009]{waterhouse2009jalview}
	Waterhouse, A.~M., Procter, J.~B., Martin, D.~M., Clamp, M., and Barton, G.~J.
	(2009).
	\newblock Jalview version 2—a multiple sequence alignment editor and analysis
	workbench.
	\newblock {\em Bioinformatics}, 25(9):1189--1191.
	
	\bibitem[Williams et~al., 2021]{williams2021molecular}
	Williams, J.~K., Wang, B., Sam, A., Hoop, C.~L., Case, D.~A., and Baum, J.
	(2021).
	\newblock Molecular dynamics analysis of a flexible loop at the binding
	interface of the SARS-CoV-2 spike protein receptor-binding domain.
	\newblock {\em Proteins: Structure, Function, and Bioinformatics}, in press.
	
	\bibitem[Wong, 2020]{wong2020assessing}
	Wong, S.~W. (2020).
	\newblock Assessing the impacts of mutations to the structure of COVID-19 spike
	protein via sequential monte carlo.
	\newblock {\em Journal of Data Science}, 18(3):511--525.
	
	\bibitem[Wong et~al., 2017]{wong2017fast}
	Wong, S.~W., Liu, J.~S., and Kou, S. (2017).
	\newblock Fast de novo discovery of low-energy protein loop conformations.
	\newblock {\em Proteins: Structure, Function, and Bioinformatics},
	85(8):1402--1412.
	
	\bibitem[Wrapp et~al., 2020]{wrapp2020cryo}
	Wrapp, D., Wang, N., Corbett, K.~S., Goldsmith, J.~A., Hsieh, C.-L., Abiona,
	O., Graham, B.~S., and McLellan, J.~S. (2020).
	\newblock Cryo-EM structure of the 2019-ncov spike in the prefusion
	conformation.
	\newblock {\em Science}, 367(6483):1260--1263.
	
	\bibitem[Wrobel et~al., 2020]{wrobel2020sars}
	Wrobel, A.~G., Benton, D.~J., Xu, P., Roustan, C., Martin, S.~R., Rosenthal,
	P.~B., Skehel, J.~J., and Gamblin, S.~J. (2020).
	\newblock SARS-CoV-2 and bat ratg13 spike glycoprotein structures inform on
	virus evolution and furin-cleavage effects.
	\newblock {\em Nature structural \& molecular biology}, 27(8):763--767.
	
	\bibitem[Yan et~al., 2020]{yan2020structural}
	Yan, R., Zhang, Y., Li, Y., Xia, L., Guo, Y., and Zhou, Q. (2020).
	\newblock Structural basis for the recognition of SARS-CoV-2 by full-length
	human ACE2.
	\newblock {\em Science}, 367(6485):1444--1448.
	
	\bibitem[Yurkovetskiy et~al., 2020]{yurkovetskiy2020structural}
	Yurkovetskiy, L., Wang, X., Pascal, K.~E., Tomkins-Tinch, C., Nyalile, T.~P.,
	Wang, Y., Baum, A., Diehl, W.~E., Dauphin, A., Carbone, C., et~al. (2020).
	\newblock Structural and functional analysis of the D614G SARS-CoV-2 spike
	protein variant.
	\newblock {\em Cell}, 183(3):739--751.
	
	\bibitem[Zhang et~al., 2021]{zhang2021structural}
	Zhang, J., Cai, Y., Xiao, T., Lu, J., Peng, H., Sterling, S.~M., Walsh, R.~M.,
	Rits-Volloch, S., Zhu, H., Woosley, A.~N., et~al. (2021).
	\newblock Structural impact on SARS-CoV-2 spike protein by D614G substitution.
	\newblock {\em Science}, 372(6541), 525-530.
	
	\bibitem[Zhang et~al., 2020]{zhang2020sars}
	Zhang, L., Jackson, C.~B., Mou, H., Ojha, A., Peng, H., Quinlan, B.~D.,
	Rangarajan, E.~S., Pan, A., Vanderheiden, A., Suthar, M.~S., et~al. (2020).
	\newblock SARS-CoV-2 spike-protein D614G mutation increases virion spike
	density and infectivity.
	\newblock {\em Nature communications}, 11, 6013.
	
	\bibitem[Zhu et~al., 2020]{zhu2020novel}
	Zhu, N., Zhang, D., Wang, W., Li, X., Yang, B., Song, J., Zhao, X., Huang, B.,
	Shi, W., Lu, R., et~al. (2020).
	\newblock A novel coronavirus from patients with pneumonia in China, 2019.
	\newblock {\em New England journal of medicine}, 382:727--733.
	
\end{thebibliography}
\end{document}